\documentclass[useAMS,usenatbib]{mn2e}
\pdfoutput=1
\usepackage{times}
\usepackage{aas_macros}
\usepackage{graphicx}

\newcommand\mdot{$M_{\odot}~\mathrm{yr^{-1}}$}

\title[Unstable colliding winds at high resolution]{High resolution numerical simulations of unstable colliding stellar winds}
\author[A. Lamberts, S. Fromang, G. Dubus]{A. Lamberts$^{1}$\thanks{E-mail:
Astrid.Lamberts@obs.ujf-grenoble.fr}, S. Fromang$^{2}$ and G. Dubus$^{1}$\\
$^{1}$UJF-Grenoble 1 / CNRS-INSU, Institut de Plan\'etologie et d'Astrophysique de Grenoble (IPAG) UMR 5274, Grenoble, F-38041, France.\\
$^{2}$Laboratoire AIM, CEA/DSM - CNRS - Universit\'e Paris 7, Irfu/Service d'Astrophysique, CEA-Saclay, 91191 Gif-sur-Yvette, France}
\begin{document}

\date{Accepted 2011 August 18. Received 2011 May 20}

\pagerange{\pageref{firstpage}--\pageref{lastpage}} \pubyear{2011}

\maketitle

\label{firstpage}

\begin{abstract}
We investigate the hydrodynamics of the interaction of two supersonic winds in binary systems. The collision of the winds creates two shocks separated by a contact discontinuity. The overall structure depends on the momentum flux ratio $\eta$ of the winds. We use the code RAMSES with adaptive mesh refinement to study the shock structure up to smaller values of $\eta$, higher spatial resolution and greater spatial scales than have been previously achieved. 2D and 3D simulations, neglecting orbital motion, are compared to widely-used analytic results and their applicability is discussed. In the adiabatic limit, velocity shear at the contact discontinuity triggers the Kelvin-Helmholtz instability. We quantify the amplitude of the resulting fluctuations and find that they can be significant even with a modest initial shear.  Using an isothermal equation of state leads to the development of thin shell instabilities. The initial evolution and growth rates enables us to formally identify the non-linear thin shell instability (NTSI) close to the binary axis. Some analogue of the transverse acceleration instability is present further away. The NTSI produces large amplitude fluctuations and dominates the long-term behaviour. We point out the computational cost of properly following these instabilities. Our study provides a basic framework to which the results of more complex simulations, including additional physical effects, can be compared.

\end{abstract}


\section{Introduction}
The stellar winds of massive stars are driven by radiation pressure to highly supersonic terminal velocities $v_{\infty}\approx 1000-3000$~km~s$^{-1}$, with mass loss rates that can reach $\dot{M}\approx 10^{-6}$ M$_\odot$~yr$^{-1}$ in O stars and $10^{-4}$ M$_\odot$~yr$^{-1}$ in Wolf-Rayet stars \citep{2008A&ARv..16..209P}. 
The interaction of two such stellar winds in a binary system creates a double shock structure where the material is condensed, heated and mixed with important observational consequences (see \citealt{2005xrrc.procE2.01P} for a review). For instance, these colliding wind binaries (CWB) have much larger X-ray luminosities than seen in isolated massive stars due to the additional emission from the shock-heated material. The increased density in the shock region also has an impact on the absorption of light within the binary. Further away from the system, free-free emission is detected in the radio, possibly supplemented by synchrotron radiation from electrons accelerated at the shock. High-resolution imaging in infrared \citep{1999Natur.398..487T} and radio \citep{2003A&A...409..217D} has made it possible to trace the large scale spiral structure created by the winds with the orbital motion of the stars. The interpretation of these observations requires knowledge of the shock structure and geometry.

Assuming a purely hydrodynamical description, the interaction results in the formation of two shocks separated by a contact discontinuity. In the adiabatic limit, the gas behind the shock is heated to temperatures $T\sim {\cal M}^2 T_w$ (where $T_w$ is the wind temperature and ${\cal M}>1$ is the Mach number of the wind). The structure is shaped primarily by the momentum flux ratio of the winds \citep{1990FlDy...25..629L}
\begin{equation}\label{eq:eta}
\eta\equiv \frac{\dot{M}_2v_{\infty 2}}{\dot{M}_1v_{\infty 1}}.
\end{equation}
The subscript 1 stands for the star with the stronger wind, the subscript 2 for the star with the weaker wind.
For reasons of symmetry, the contact discontinuity is on the midplane between the stars when $\eta=1$. \citet{Pilyugin:2007rx} obtained a complete semi-analytic description of the interaction region for this specific case. When $\eta\neq1$ the shock structure bends towards one of the stars as the stronger wind gradually overwhelms the weaker wind.  This leads to a bow shock shape close to the binary and the contact discontinuity shows an asymptotic opening angle at large scales \citep[neglecting orbital motion,][]{1987A&A...183..247G}. The shock structure must then be derived from numerical simulations \citep{Luo:1990mp}. It depends on other parameters (Mach number, velocity ratio of the winds) and, crucially, on the cooling properties of the gas. Cooling becomes efficient when the radiative time scale of the shocked flow becomes shorter than its dynamical time scale \citep{Stevens:1992on}. In this case, the kinetic energy of the wind (typically $\sim 10^{36}$ erg~s$^{-1}$) is radiated away and the incoming gas is strongly decelerated at the shock ($v= v_{\infty}/{\cal M}^2$ in the isothermal limit compared to $v=v_{\infty}/4$ in the adiabatic limit). The interaction region becomes thin and the double shock structure indistinguishable from the contact discontinuity. Analytical solutions for the interaction geometry can be derived in the limit of an infinitely thin shock (\citealt{1987A&A...183..247G,Luo:1990mp,1993MNRAS.261..430D,Canto:1996jj,2009ApJ...703...89G}, see \S3 below). 

The analytical solutions provide useful approximations but their validity may be questioned as numerical simulations show that shocks become unstable (see \S4). The contact discontinuity separates two media with different tangential velocities, triggering the Kelvin-Helmholtz instability (KHI) in adiabatic or radiatively-inefficient shocks. The impact is more or less pronounced \citep{Stevens:1992on,Lemaster:2007sl,2009MNRAS.396.1743P,2010MNRAS.406.2373P,2011A&A...527A...3V} and has not been quantified yet. Thin shocks become violently unstable and have garnered more attention. The instability was initially seen in simulations where the gas was assumed to be isothermal, mimicking the effect of efficient cooling (\citealt{Stevens:1992on,1998NewA....3..571B} but see \citealt{1998MNRAS.298.1021M}), and has since also been seen in simulations including a more realistic  treatment of radiative cooling  \citep{2009MNRAS.396.1743P,2011A&A...527A...3V}. The resulting mixing and variability can have important observational consequences. The origin of the instability remains unclear \citep{1998Ap&SS.260..215W}. Two mechanisms have been proposed in the thin shell limit: the {non-linear thin shell instability} (NTSI, \citealt{1994ApJ...428..186V}) and the {transverse acceleration instability} (TAI, \citealt{1993A&A...267..155D,1996ApJ...461..927D}) ; both may be at work in colliding winds \citep{1998NewA....3..571B}.

Much progress has been made in including more realistic physics in simulations of CWB (wind acceleration, gravity from the stars, radiative inhibition, cooling functions, heat conduction, orbital motion etc.). These are undoubtedly important effects to consider when comparing with observations but they complicate the comparison with basic analytical expectations which, in turn, makes it more difficult to assess their contributions. Here, we present simulations neglecting all these effects, assuming a polytropic gas $P\propto \rho^\gamma$ with $\gamma=5/3$ (adiabatic) or  $\gamma=1$ (isothermal). Our purpose is to understand how the shock region compares to expectations and to constrain the conditions giving rise to instabilities particularly in the limit of low $\eta$. We performed a systematic set of 2D and 3D numerical simulations using the hydrodynamical code RAMSES~\citep{2002A&A...385..337T} with adaptive mesh refinement, allowing us to reach the high resolutions required for thin shocks and low $\eta$ while keeping a wide simulation domain to study the asymptotic behaviour  (\S2). Notable features of the wind interaction region are discussed and compared to the analytical solutions: shock location, width, opening angle and the presence of reconfinement shocks at low $\eta$ (\S3). We present our investigations of the instabilities in the adiabatic and isothermal case in \S4. We find that the non linear thin shell instability (NTSI) is the dominant mechanism for isothermal winds. We then replace this work in its larger context, discussing the impact that including additional physics would have on our conclusions and the computational cost required to follow the instabilities (\S5).

\section{Numerical Simulations}

We use the hydrodynamical code RAMSES (\citealt{2002A&A...385..337T}) to perform our simulations. This code uses a second order Godunov method to solve the equations of hydrodynamics
\begin{eqnarray}
	\frac{\partial\rho}{\partial t}+\nabla \cdot(\rho \mathbf{v})    &=&  0\\  
	\frac{\partial(\rho \mathbf{v})}{\partial t}+\nabla \cdot (\rho \mathbf{v}\mathbf{v})+\nabla P 	&=& 0	\\
	\frac{\partial E }{\partial t}+\nabla \cdot[\mathbf{v}(E+P)]	&=& 0
\end{eqnarray}
where $\rho$ is the density, $\mathbf{v}$ the velocity and $P$ the pressure of the gas. The total energy density $E$ is given by
\begin{equation}\nonumber
E= \frac{1}{2}\rho v^2+\frac{P}{(\gamma -1)}
\end{equation}
$\gamma$ is the adiabatic index, its value is 5/3 for adiabatic gases and 1 for isothermal gases. For numerical reasons $\gamma$ is set to 1.01 for isothermal simulations~\citep{1998ApJ...495..821T}. We use the MinMod slope limiter. We compare our simulations with analytic solutions in \S3.  In order to do this, we prevent the development of instabilities in the shocked region by using the local Lax-Friedrich Riemann solver, which is more diffusive. An exact Riemann solver is used to study the development of instabilities in \S\ref{instabilities}. We perform 2D and 3D simulations on a Cartesian grid with outflow boundary conditions. We use adaptive mesh refinement (AMR) which enables to locally increase the spatial resolution according to the properties of the flow. In 2D the grid is defined by a coarse resolution $n_x=128$ with up to 6 levels of refinement. In 3D the grid is defined by $n_x=32$ with up to 5 levels of refinement. The refinement criterion is based on density gradients.

\subsection{Model for the winds} 
Our method to implement the winds is similar to the one developed by \citet{Lemaster:2007sl} and described in the appendix of their paper. The main aspects are recalled here for completeness. Around each star, we create a wind by imposing a given density, pressure and velocity profile in a spherical zone called mask. The masks are reset to their initial values at all time steps to create steady winds. The velocity is purely radial and set to the terminal velocity $v_{\infty}$ of the wind  in the whole mask. Setting the velocity to $v_{\infty}$ supposes the winds have reached their terminal velocity at the interaction zone. This might not be applicable for very close binaries or if $\eta\ll 1$ because the shocks are then very close to one of the stars. Our 2D setup differs from those usually found in the literature (e.g.~\citealt{Stevens:1992on,1995MNRAS.277...53B,2006A&A...446.1001P}) in that we work in the cylindrical $(r,\theta)$ plane instead of the $(r,z)$ plane. A drawback of our 2D method is that the structure of the colliding wind binary is not identical when going from a 2D to 3D simulation with the same wind parameters. However, as described later, we found that the 3D structure is mostly recovered in 2D by using the scaling $\sqrt{\eta_{\rm 3D}} \rightarrow \eta_{\rm 2D}$. An advantage of our 2D approach is that it is straightforward to include binary motion without resorting to full 3D simulations. Such simulations will be described elsewhere~(see \citealt{lamberts} for preliminary calculations).
 The density profile is determined by mass conservation through the mask 

\begin{equation}\label{eq:rho}
\rho_{3D} = \frac{\dot{M}}{4\pi r^2 v_{\infty}}  \hskip0.35cm \textrm{ (3D)} \hskip 0.65cm\rho_{2D} = \frac{\dot{M}}{2\pi r v_{\infty}}\hskip.35cm\textrm{(2D)}
\end{equation}
where $r$ is the distance to the centre of the mask. The pressure is determined using $P\rho^{-\gamma}=K$ with $K$ constant in each region. 
Time is expressed in years and mass loss rates are expressed in $10^{-8}$\mdot. We decide to scale all distances to the binary separation $a$. This way the results of a simulation can easily be rescaled to systems with a different separation.  For each simulation, the input parameters are the mass loss rate, terminal velocity and Mach number $\mathcal{M}$ at $r=a$ of each wind. We then derive the hydrodynamical variables at $a$. After that the corresponding density, pressure and velocity profile in the mask are computed.

For $\eta \ll 1$ the shocks form very close to the second star. In this case, the mask of the star has to be as small as possible so that the shocks can form properly \citep{1998MNRAS.300..479P}. However a minimum length of 8 computational cells per direction is needed to obtain spherical symmetry of the winds. We thus fix the size of the masks to 8 computational cells in each direction for the highest value of refinement. We performed tests with a single star for different sizes of the mask ranging from 0.03$a$ to 1.5$a$. The tests were performed for $n_x=128$ and 4 levels of refinement. The resulting density profiles all agree with the analytic solution with less than 1 $\%$ offset. The surrounding medium is filled with a density $\rho_{amb}=10^{-4}\rho(a)$ and pressure $P_{amb}=0.1P(a)$. This initial medium is pushed away by the winds. Simulations with different $\rho _{amb}$ and $P_{amb}$ show the same final result, to round-off precision. The size of the computational domain varies between $l_{box}=2a$ and $l_{box}=80a$ according to the purpose of the simulation. Except where stated otherwise, we took $\dot{M_1}=\dot{M_2}=10^{-7}$ \mdot, $\mathcal{M}_1=\mathcal{M}_2=30$, $v_{\infty 2}=2000$\, km\, s$^{-1}$ and $\eta$ was varied by changing $v_{\infty 1}$.Hence, our low momentum flux ratios can imply very high velocities for the first wind.

\section{The shock region}\label{shock}
In this section we study the dependence on $\eta$ of the geometry of the interaction zone. We discuss he analytic solutions for the colliding wind geometry, in 2D and 3D, to which we compare our simulations. Simulations are performed with adiabatic and isothermal equations of state. In both cases the numerical diffusion introduced by the solver is sufficient to quench the development of instabilities. Section~\ref{instabilities} deals with high resolution simulations of the development of these instabilities. 

\subsection{Analytical approximations\label{analytical}}
\begin{figure}
\centering
\includegraphics[width = .4\textwidth]{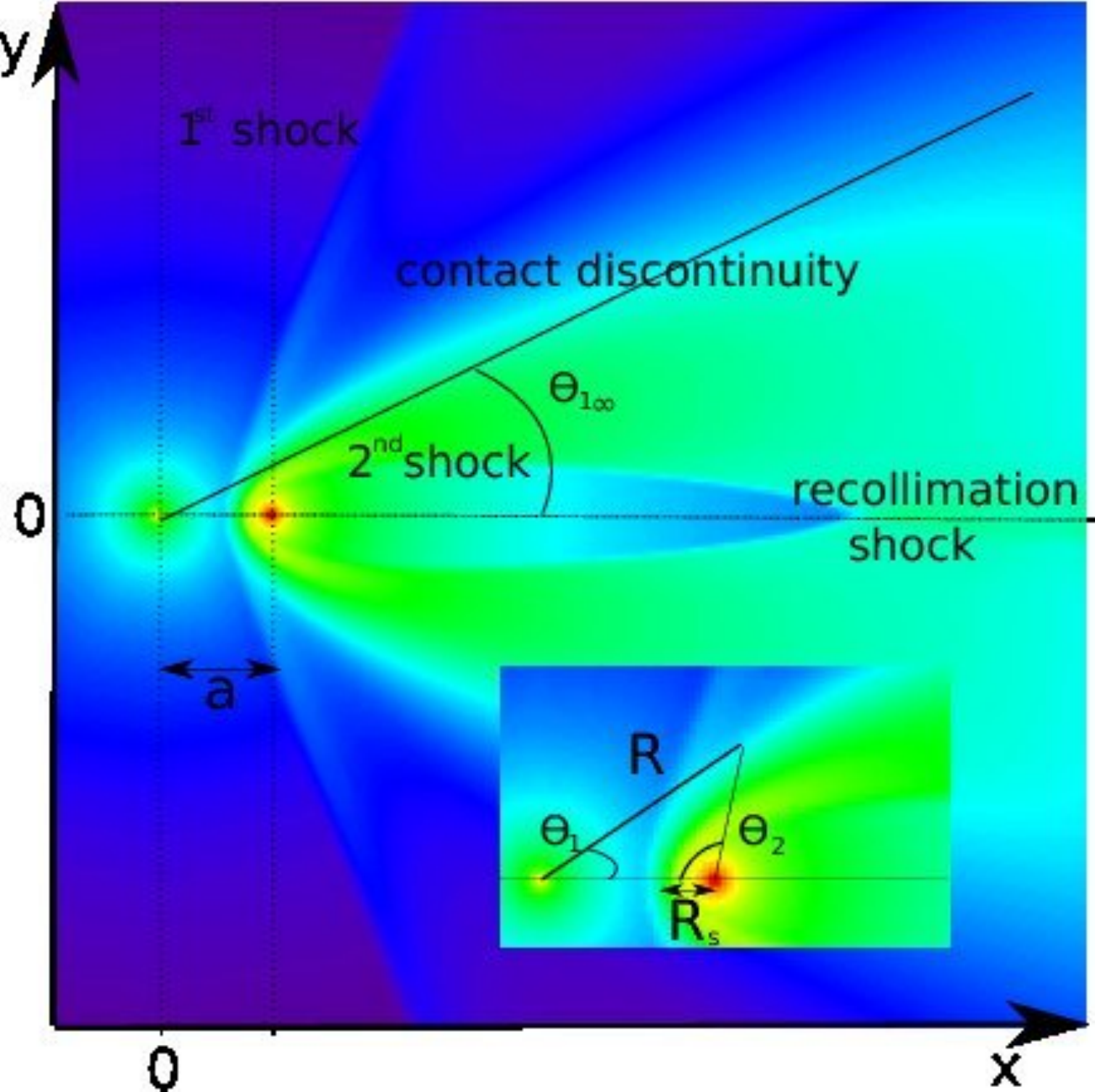}
\caption{ Density map of the interaction zone for $\eta=1/32=0.03125$ (3D simulation). It is a cut perpendicular to the line of centres taken from a 3D simulation. A zoom on the binary system is shown at the bottom right corner. The stars are positioned at the intersections of the dotted lines. The first star has coordinates (0,0), the second one has coordinates ($a$,0). There are three density jumps (for increasing $x$). The first shock separates the unshocked wind from the first star from the shocked wind. The contact discontinuity separates the shocked winds from both stars. It intersects the line of centres at the standoff point $R_s$. The second shock separates the shocked and unshocked parts of the wind from the second star. As the wind from the second star is collimated, there is a reconfinement shock along the line of centres. $R(\theta_1$) is the distance between the contact discontinuity and the first star, $\theta_1$ is the polar angle. The asymptotic opening angle is given by $\theta_{1\infty}$. $l$ is the distance to $R_s$ along the contact discontinuity.}
\label{fig:geometry} 
\end{figure}

The overall structure of the colliding wind binary is given in Fig.~\ref{fig:geometry}. The density map shows two shocks separating the free winds from the shocked winds. The shocked winds from both stars are separated by a contact discontinuity. The Bernouilli relation is preserved across shocks hence
\begin{equation}
\frac{1}{2}v_{\infty 1}^2=\frac{\gamma}{\gamma-1}\frac{P_{1s}}{\rho_{1s}}+\frac{1}{2}v_{1s}^2
\label{eq:Bernouilli}
\end{equation}
across the first shock. The subscript $s$ refers to quantities in the shocked region and we have neglected the thermal pressure in the unshocked wind due to its high Mach number. A similar equation holds for the second shock. The Bernouilli relation is constant in each shocked region but discontinuous at the CD. There, $P_{1s}\equiv P_{2s}$ by definition and $v_{1s}=v_{2s}=0$ on the line-of-centres so that the two Bernouilli equations combine to give $\rho_{1s}v_{\infty 1}^2=\rho_{2s}v_{\infty 2}^2$, with  $\rho_s$ the value of the density on each side of the contact discontinuity. Assuming that the density is constant in each shocked region on the binary axis (the numerical simulations carried out below show this is a very good approximation) then  
\begin{equation}
\rho_{1} v_{\infty 1}^2\approx \rho_{2}v_{\infty 2}^2
\end{equation}
where $\rho_{1}$ ($\rho_2$) is the value of the density at the first (second) shock.
The above relation states the balance of ram pressures \citep{Stevens:1992on}. Using Eqs.~(\ref{eq:eta}) and~(\ref{eq:rho}) then yields 
\begin{equation}\label{eq:r1_r2}
r_2\approx \sqrt{\eta} r_1 \hskip0.35cm \textrm{ (3D)} \hskip 0.65cm r_2\approx \eta r_1 \hskip0.35cm\textrm{ (2D)}
\end{equation}
where $r_1$ is the distance between the first star and the first shock and $r_2$, the distance between the second star and the second shock. If the shock is thin then $r_1+r_2\approx a$ and the distance $R_s\approx r_2$ of the CD to the second star is 
\begin{equation}\label{standoff}
\frac{R_s}{a}\approx \frac{\sqrt{\eta}}{1+\sqrt{\eta}} \hskip0.35cm \textrm{ (3D)} \hskip 0.65cm\frac{R_s}{a}\approx \frac{\eta}{1+\eta} \hskip0.35cm\textrm{ (2D)}
\end{equation}
Note that, for a given $\eta \leq 1$, the contact discontinuity is closer to the second star for a 2D geometry than for a 3D geometry.

The shock positions are not easily derived away from the line-of-centres, where the density is not constant in the shocked winds. Analytic solutions have been derived based on the thin shell hypothesis, which considers both shocks and the contact discontinuity are merged into one single layer. \citet{Stevens:1992on} (see also \citealt{Luo:1990mp}, \citealt{1993MNRAS.261..430D} and \citealt{Antokhin:2004hi}) derive the following equation for the shape of the interaction region by assuming that it is located where the ram pressures normal to the shell balance:
\begin{equation}\label{antokhin_3d}
 \frac{{ d}x}{{ d}y}=\frac{x}{y} - \left(\frac{a}{y}\right) \left[1+ \sqrt{\eta}\left(\frac{r_2}{r_{1}}\right)^{2}\right]^{-1}
\end{equation}
The same analysis for the 2D structure (Eq.~\ref{eq:rho}) leads to
\begin{equation}
\label{stevens2d}
 \frac{{ d}x}{{ d}y}=\frac{x}{y} - \left(\frac{a}{y}\right) \left[1+ \sqrt{\eta}\left(\frac{r_2}{r_{1}}\right)^{3/2}\right]^{-1}
\end{equation}
\citet{Canto:1996jj}, extending the work of  \cite{Wilkin:1996ud}, found an analytical solution in the thin shell limit based on momentum conservation (hence, taking into account the centrifugal correction {\em i.e.} the forces exerted on the gas as it follows a non-linear path along the shock, \citealt{1971SPhD...15..791B,1975Ap&SS..35..299D,1987A&A...183..247G}):
\begin{equation}\label{canto_3d}
 {\theta_2}{\cot \theta_2}-1=\eta\left({\theta_1}{\cot\theta_1}-1\right)
\end{equation}
(see Fig.~\ref{fig:geometry} for the definition of $\theta_1$ and $\theta_2$.) The same analysis in 2D leads to 
\begin{equation}
\label{canto2d}
\frac{\cos\theta_2-1}{\sin \theta_2}=\eta \frac{\cos\theta_1-1}{\sin \theta_1}
\end{equation}

\subsection{2D study\label{2dstudy}}
\begin{figure}
\centering
\includegraphics[width = .48\textwidth]{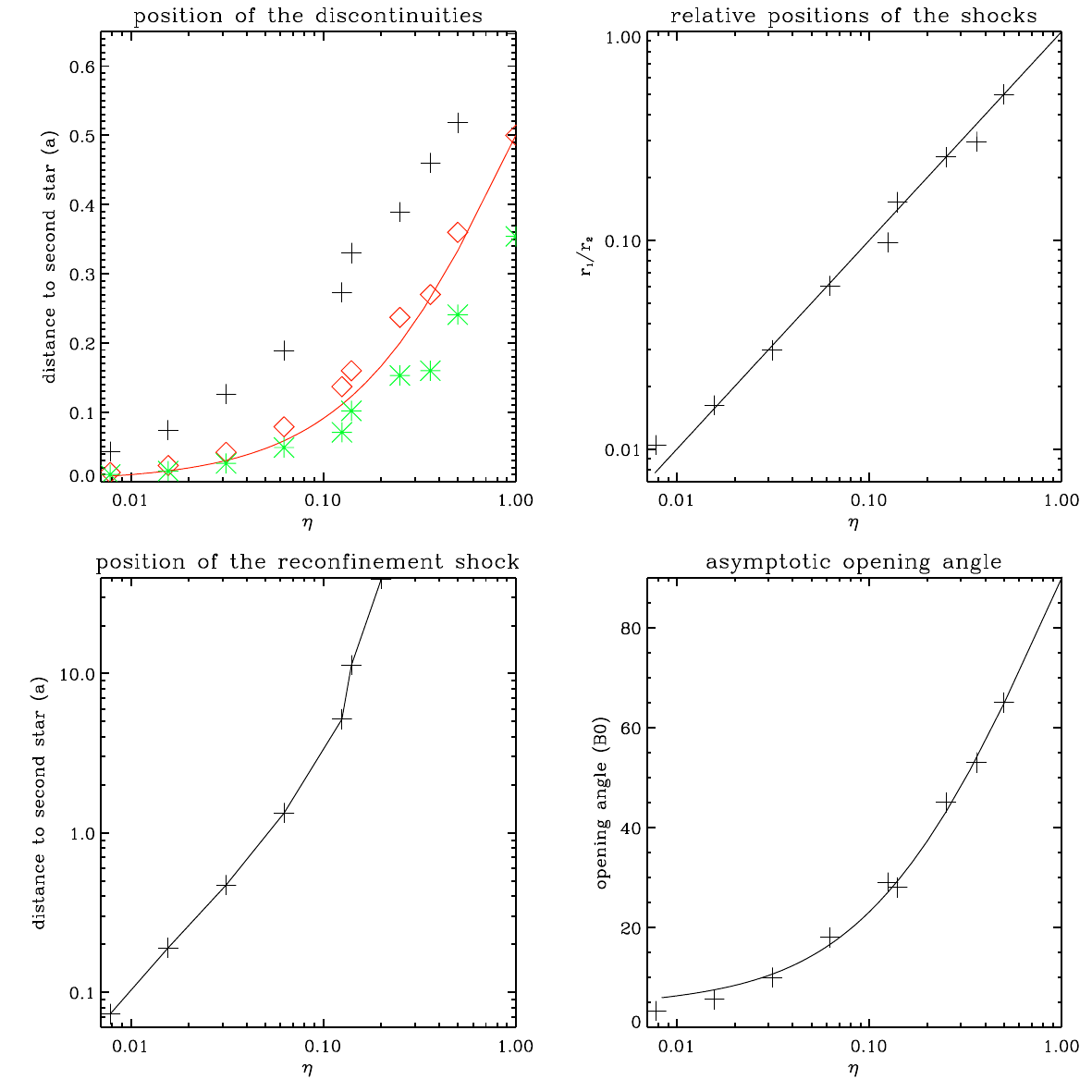}
\caption{Dependence of the shock geometry with $\eta$ in 2D. Top left panel: Position of the different density jumps: first shock (black crosses), contact discontinuity (blue diamonds) and second shock (green asterisks). The 2D analytic solution for the contact discontinuity (Eq.~\ref{standoff}) is overplotted (blue solid line). Top right panel: ratio of the shock positions measured from the simulations and compared to Eq.~\ref{eq:r1_r2}. Bottom left panel : position of the reconfinement shock. Bottom right panel: asymptotic opening angle (crosses) compared with the asymptotic angle derived from the \citet{Canto:1996jj} (dashed line) and   \citet{Stevens:1992on} (solid line) solutions.}
\label{fig:shock_pos} 
\end{figure}

We performed a systematic study of the 2D geometry of the interaction zone in the adiabatic case for $\eta$ ranging from 1 down to $1/128$ with Mach number ${\cal M}=30$ for both winds. Fig.~\ref{fig:shock_pos} shows how the main features of the colliding wind binary vary with $\eta$. The positions of the discontinuities on the binary axis (top left) were computed by determining the local extrema of the slope of the density, excluding the masks. There is very good agreement with the analytic solution for the position of the standoff point (Eq.~\ref{standoff}). The relation for the ratio of shock positions (Eq.~\ref{eq:r1_r2}) is also verified (top right). As $\eta$ decreases both shocks and the contact discontinuity get closer to the star with the weaker wind. Since the thickness of the shell decreases as $\eta$ decreases, proper modelling of the interaction region for low $\eta$ requires a higher numerical resolution. For $\eta \la 0.25$, the second wind is totally confined and there is a reconfinement shock on the line of centres behind the second star (see Fig.~\ref{fig:geometry}). This shock draws closer to the second star as $\eta$ decreases (Fig.~\ref{fig:shock_pos}, bottom left). Similar structures were found by \citet{Myasnikov:1993wk} and \citet{2008MNRAS.387...63B} (in the latter case for $\eta<1/800$). The last panel (bottom right) shows the asymptotic opening angle of the contact discontinuity. The solution from  \citet{Stevens:1992on} gives a better agreement for low values of $\eta$.

\begin{figure*}
\centering
\includegraphics[width = .29\textwidth]{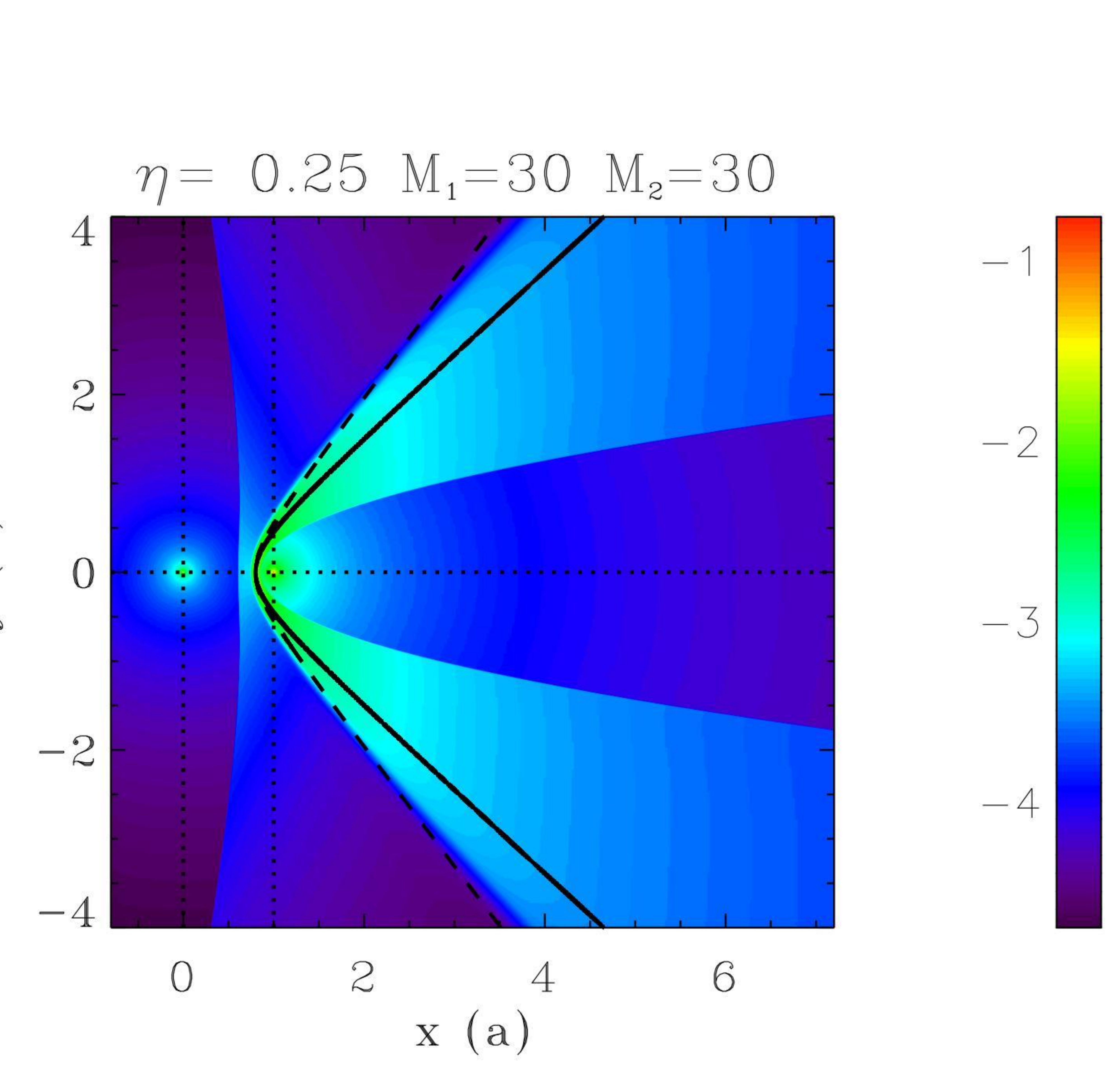}
\includegraphics[width = .29\textwidth]{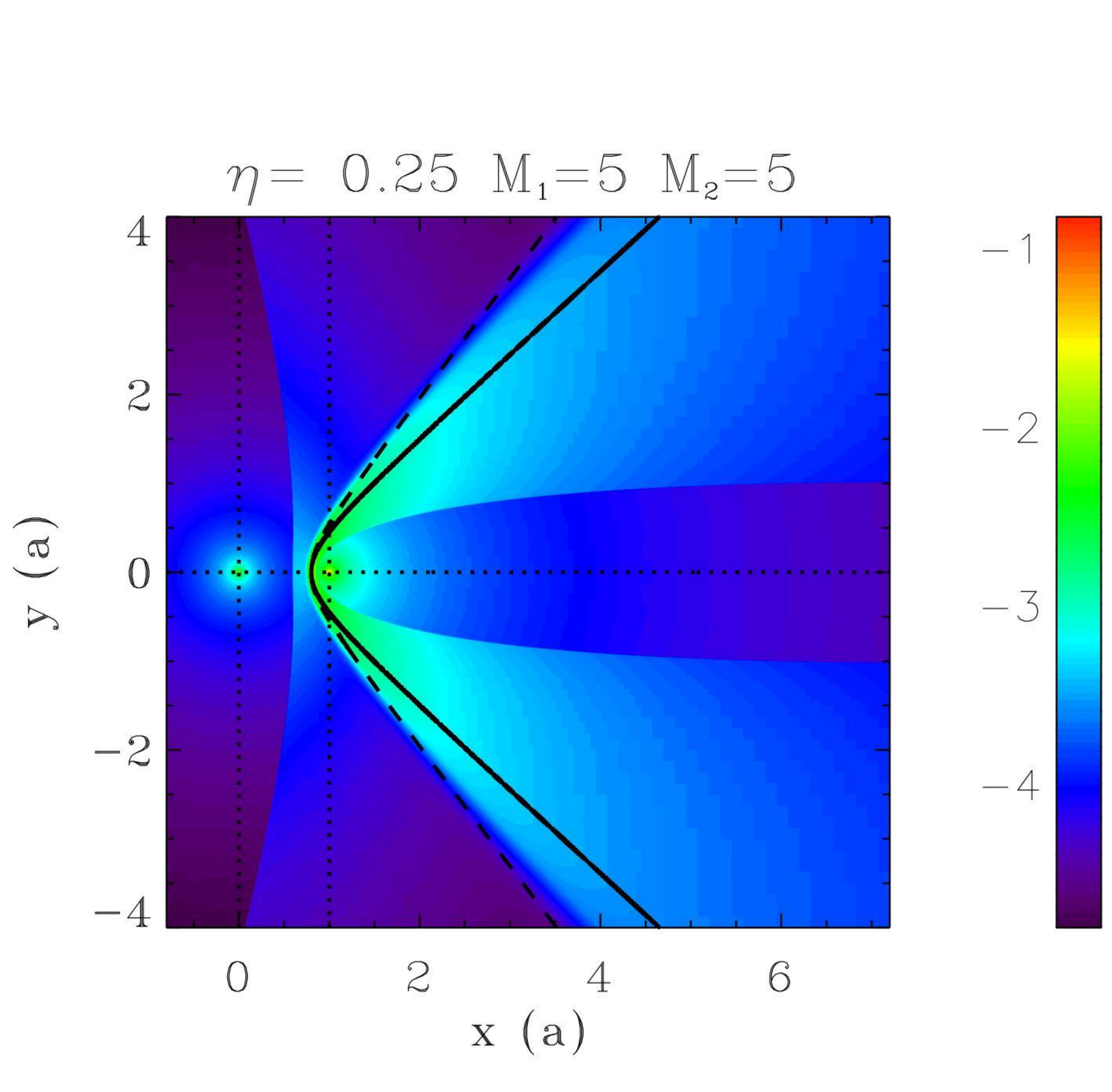}
\includegraphics[width = .29\textwidth]{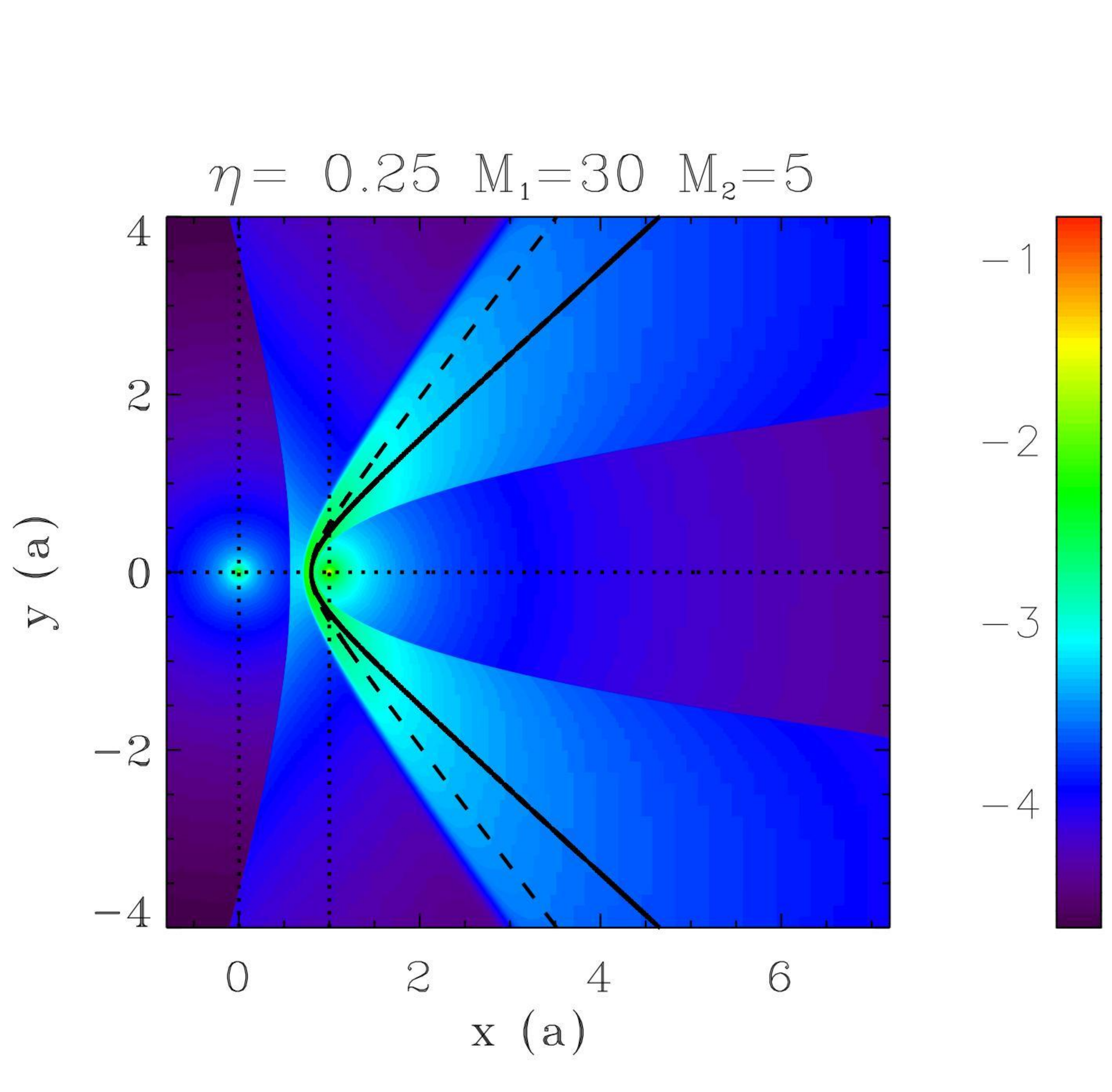}
\caption{Density maps for 2D simulations with $\eta=0.25$ and different Mach numbers for the winds (${\cal M}_1$, ${\cal M}_2$). The 2D analytic solutions derived from the assumptions of \citet{Canto:1996jj} and \citet{Stevens:1992on} are represented respectively by the dashed and solid line. The analytic solutions both assume infinite Mach numbers for both winds.}
\label{fig:Mach} 
\end{figure*}

For given Mach numbers, the geometrical structure of the colliding wind binary is set by $\eta$. We performed a series of tests for $\eta=1/8=0.125$ and different combinations for $v_{\infty 1}$, $v_{\infty 2}$, $\dot{M_1}$ and $\dot{M_2}$.  Although the density and velocity fields were different in all cases, both shocks and the contact discontinuity were located at the same place along the line of centers. Further away from the star we notice that the reconfinement shock position changes up to $\simeq 25 \%$ when changing  the velocity and mass loss rate of the winds. All other discontinuities are located at the same place. Simulations with ${\mathcal{M}_1=\mathcal{M}_2=100}$ do not show differences from the case $\mathcal{M}_1=\mathcal{M}_2=30$, as could be expected since  thermal pressure is negligible in both cases.  However, the structure for given $\eta$ depends somewhat on the Mach number of the winds if these are not very large. Fig.~\ref{fig:Mach} shows the density maps for 2D simulations with $\eta=0.25$ but with different values for the wind Mach numbers obtained by changing the wind temperature. If both winds have $\mathcal{M}=5$ instead of $\mathcal{M}=30$, the shocked region is wider and a reconfinement shock appears at $\approx 15a$ (beyond the region shown in Fig.~\ref{fig:Mach}). The position of the contact discontinuity remains the same. When both winds have different Mach numbers, the whole shocked structure is more bent towards the wind with the higher Mach number : thermal pressure from the low Mach number wind is not negligible in the shock jump conditions (see Eq.~\ref{eq:Bernouilli}) and the added term displaces the shock away from the low Mach number wind.

We also investigated the overall structure in the isothermal case, quenching the strong instabilities that are present in this case (see \S\ref{isothermal}) by using a highly diffusive solver. In this case pressure support is weaker and the shell is much thinner, as expected. The double shock structure and CD are only visible on the line of centres when using a very high spatial resolution. The position of the thin shock structure on the line of centres is within 10$\%$ of the CD position found in adiabatic simulations. The asymptotic angle is difficult to assess as the shock structure is smoother than in the adiabatic case (see e.g. Fig.~\ref{fig:3D} below) but the bracketing values are consistent with those found in the adiabatic case. We find that the weaker wind can be fully confined as in the adiabatic case. However, this occurs further away from the star than in the adiabatic case shown in Fig.~\ref{fig:shock_pos} (at $\approx 6.4a$ for $\eta=1/16$ and 2.2$a$ for $\eta=1/32$).

\subsection{3D study\label{3dstudy}}
\begin{figure*}
\centering
\includegraphics[width = .24\textwidth]{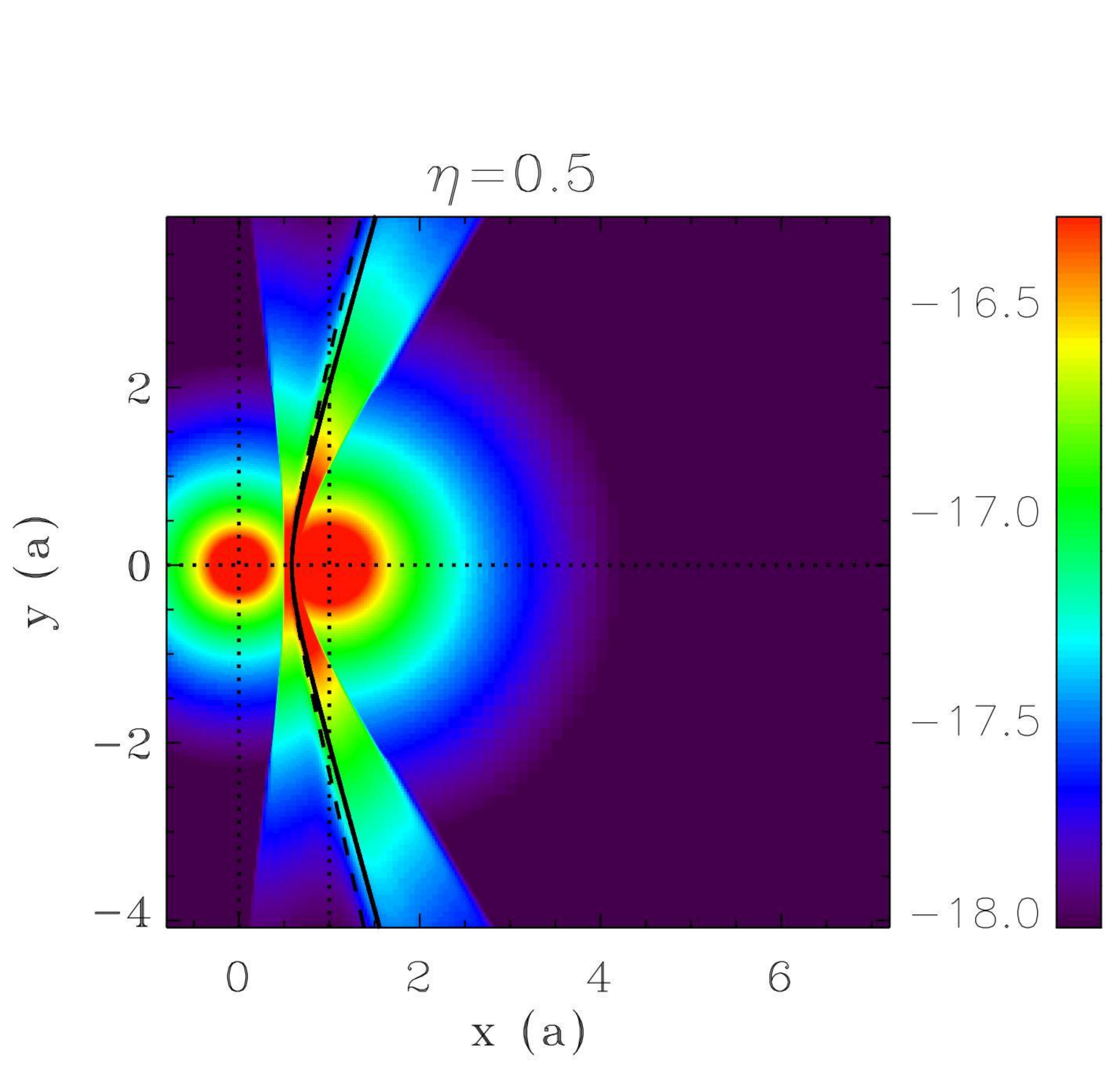}
\includegraphics[width = .24 \textwidth]{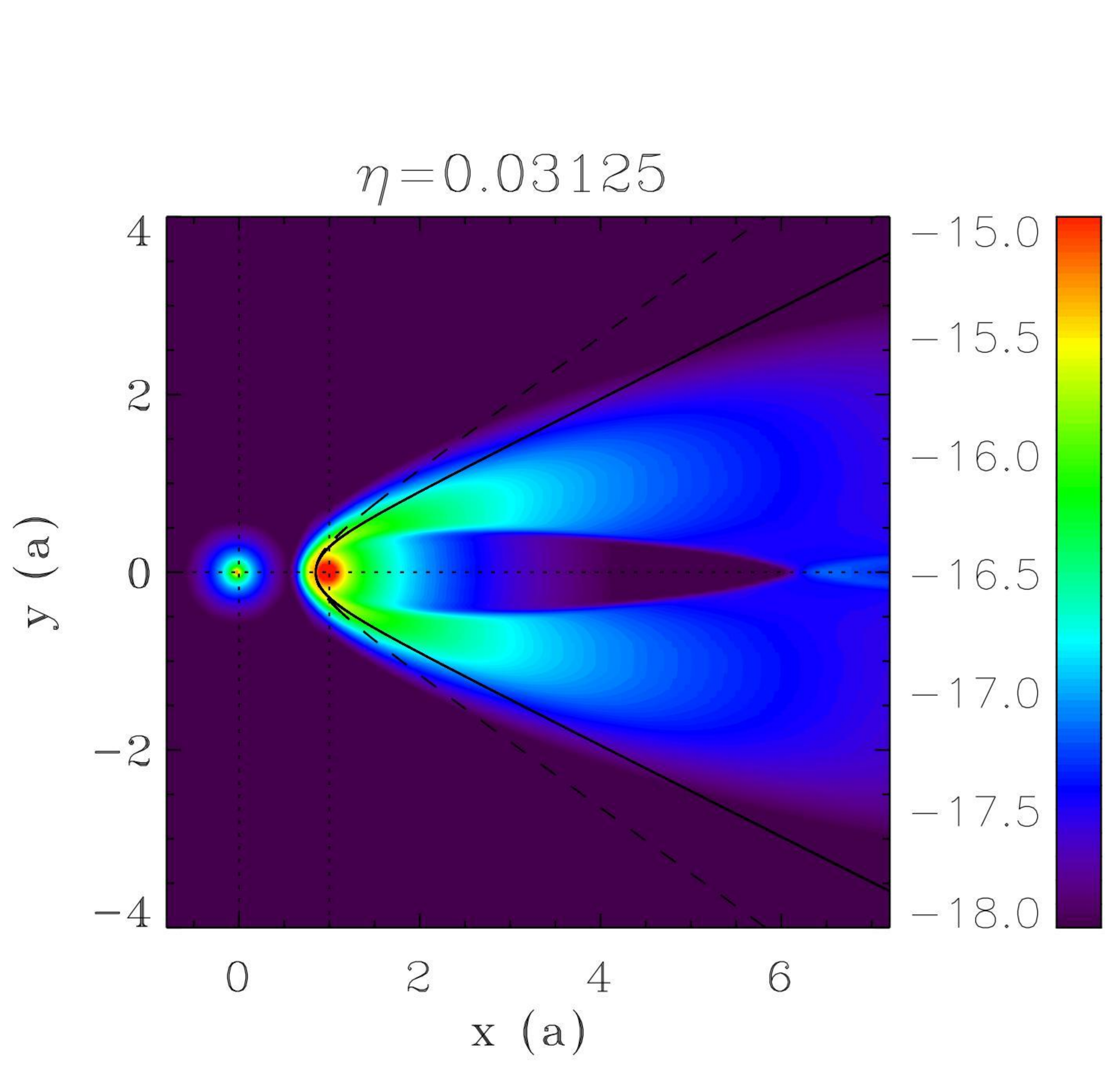}
\includegraphics[width = .24\textwidth]{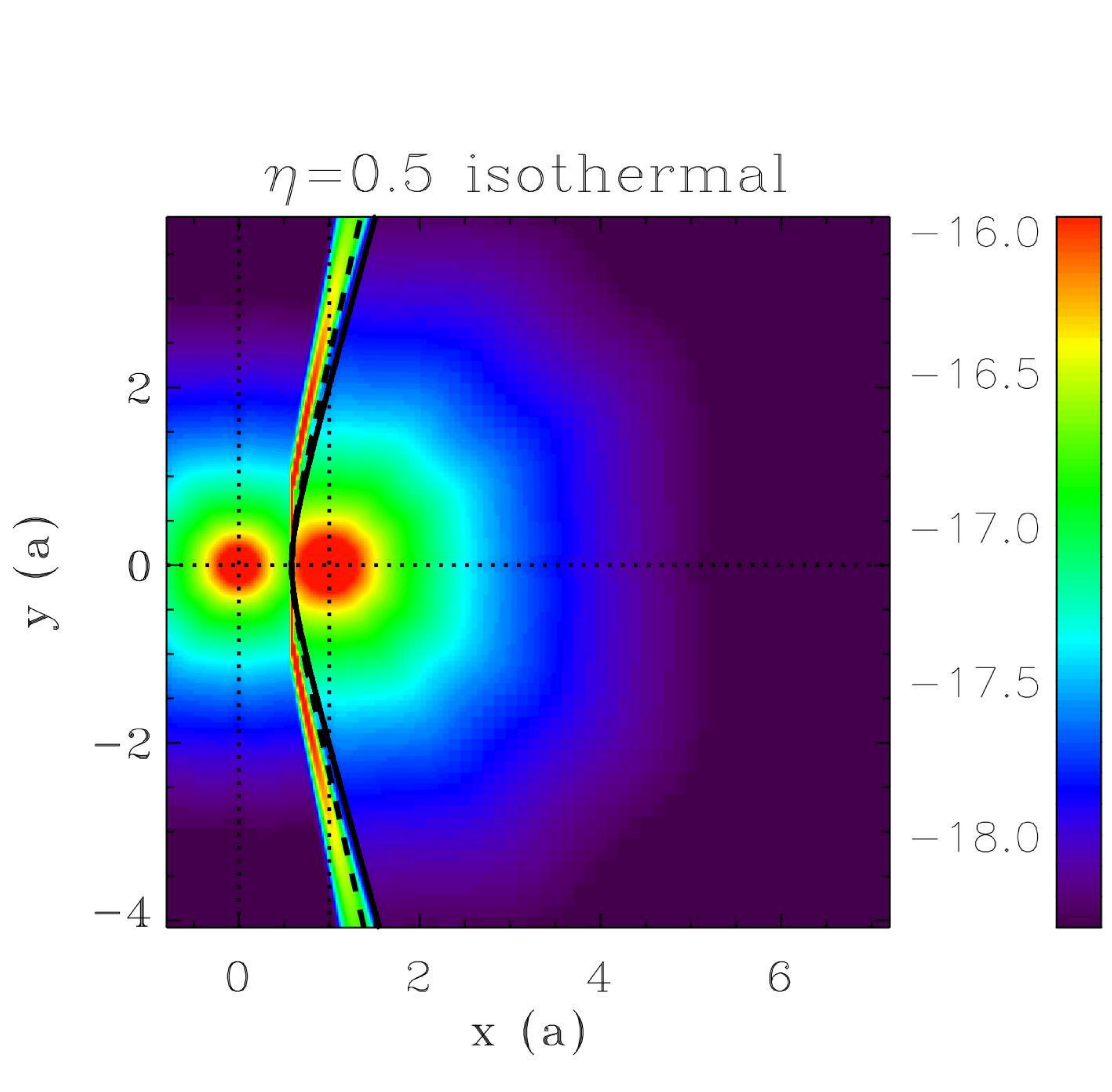}
\includegraphics[width = .24 \textwidth]{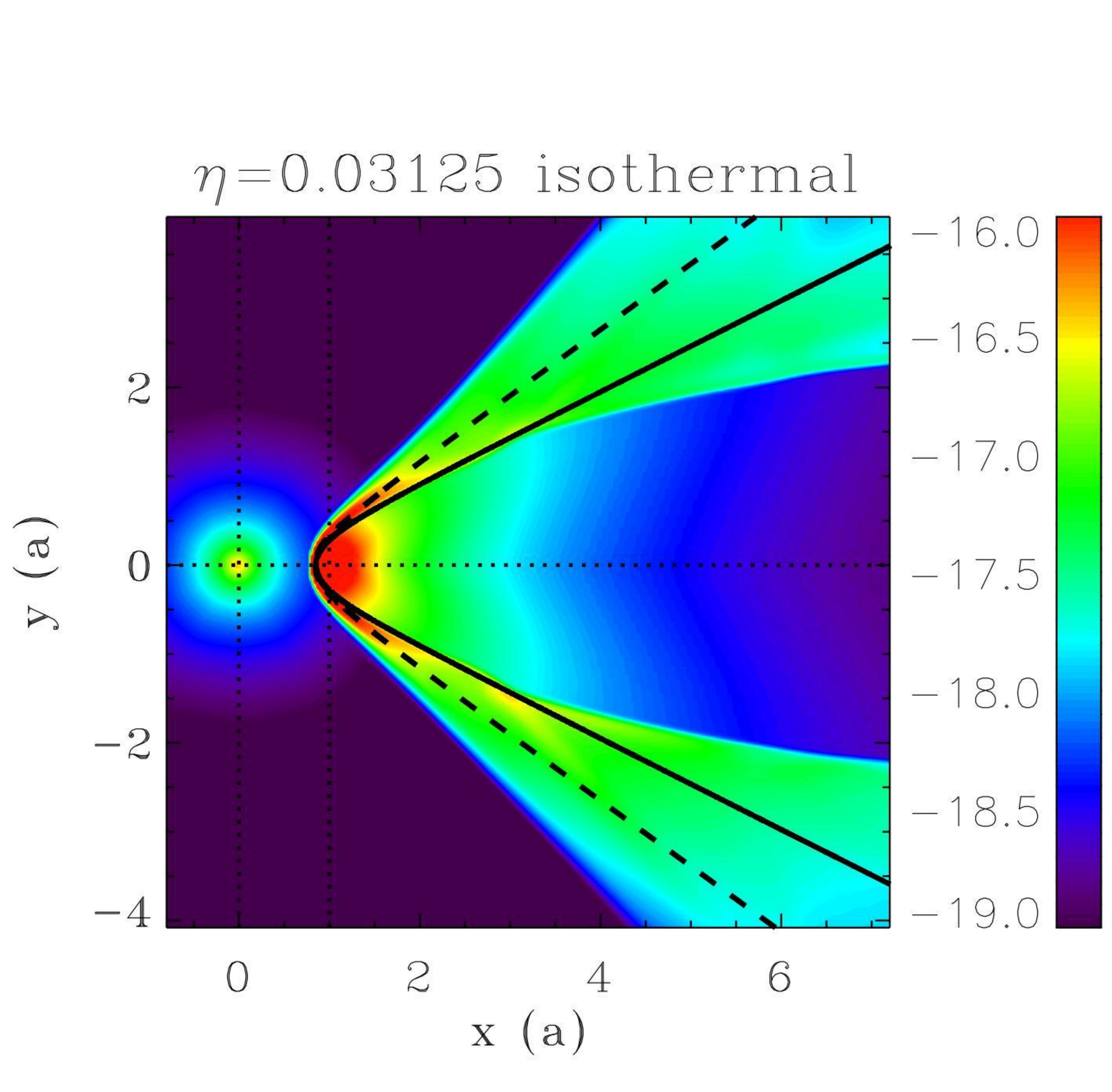}
\caption{Density maps for 3D simulations with $\eta=0.5$  and $\eta=1/32=0.03125$ in the adiabatic ($\gamma=5/3$) and isothermal ($\gamma=1.01$) limits. The stars are located at the intersections of the dotted lines. The dashed line represents the solution from \citet{Canto:1996jj}, the solid line the solution from \citet{Stevens:1992on}. The length scale is the binary separation $a$.}
\label{fig:3D} 
\end{figure*}

We completed this 2D study with the analysis of a few large scale 3D simulations, computationally more expensive than the previous 2D simulations. Fig.~\ref{fig:3D} shows the density maps for adiabatic and isothermal 3D simulations with $\eta=0.5$ and $\eta=1/32$ ($\cal M$=30). In the adiabatic case, one can clearly see the two shocks and the contact discontinuity. For $\eta=1/32$ the weaker wind is totally confined with maximum extension along the axis up to $5a$ away behind the star. For $\eta=1/64\approx 0.016$ (not shown) we find the reconfinement shock occurs at 1.0$a$. This is consistent with the 2D results  (Fig.~\ref{fig:shock_pos}) if assuming the rough mapping $\sqrt{\eta_{\rm 3D}} \rightarrow \eta_{\rm 2D}$ suggested by Eq.~\ref{standoff}. Indeed, we find no reconfinement shock for 3D simulations with $\eta=0.08$ ( which would correspond to $\eta_{\rm 2D}\approx 0.29$ in Fig.~\ref{fig:shock_pos}).\citet{2006MNRAS.372..801P} performed 2D axisymmetric simulations showing a reconfinement shock for $\\eta =0.02$ but not for $\eta=0.036$. We performed several 3D  simulations with ${\eta=1/32=0.03125}$ or $\eta=0.02$ and for different values of the Mach number $\mathcal{M}$ (assumed identical in both winds). We found that reconfinement occurs in all cases when $\mathcal{M}=30$ or 100 but that no reconfinement occurs for $\eta=0.02$ or $\eta=1/32$ when $\mathcal{M}=5$. As in the 2D case, non-negligible thermal pressure has an impact on the structure of the colliding wind binary. Whereas the presence of reconfinement for low $\eta$ and high Mach numbers around a threshold value 0.02-0.03 appears robust, the precise determination of this threshold value or of the properties of the reconfinement region is sensitive to the exact wind properties (Mach number). Radiative cooling, which is neglected here, can also have an impact on reconfinement (e.g. 2D isothermal simulation showed reconfinement further away from the star than in the adiabatic case, \S3.2).

The positions of the discontinuities  along the line of centres agree within 2$\%$ with the expected values. As with the 2D case, the shock shape is better approximated by the solution of  \citet{Stevens:1992on} at low $\eta$. For $ \eta=0.5$ we find $\theta_{\infty}=71\degr$ whereas the asymptotic angle from both \citet{Stevens:1992on} and \citet{Canto:1996jj} give 78$\degr$; for $\eta=1/32=0.03125$ we get 23$\degr$ compared to theoretical estimates of 27$\degr$  \citep{Stevens:1992on} and 35$\degr$ \citep{Canto:1996jj}. On the other hand, Figs.~\ref{fig:Mach}-\ref{fig:3D} show that the analytic solution of \citet{Canto:1996jj} is a better approximation to the contact discontinuity shape at  high $\eta$. For $\eta \geq 1/32$,  close to the line of centres, the shocked region is  thinner in the 3D case than in the 2D. For smaller values of $\eta$, the shocked zone is thicker in the 3D case. In all cases the contact discontinuity is further away from the second star in the 3D case than in the 2D case.

We have studied the geometry of the interaction region in 2D and 3D. We conclude that analytic solutions give satisfactory agreement with the results of the simulations. The solution based on ram pressure balance normal to the shock reproduces better the asymptotic opening angle of the flow at low $\eta$. We also find that the weaker wind can be entirely confined for low values of $\eta$. However, the interaction region is susceptible to instabilities that can modify these conclusions. This is investigated in the next section.

\section{Instabilities}\label{instabilities}
\subsection{The Kelvin-Helmholtz instability (KHI)\label{KH}}

When the exact Riemann solver is used, there is less numerical diffusion and the velocity shear at the contact discontinuity leads to the development of the KHI. The interface of two fluids is unstable to any velocity perturbation along the flow in the absence of surface tension or gravity \citep{1961hhs..book.....C}. The growth rate of the instability in the linear phase is $\tau_{KHI}= \lambda/(2\pi \Delta v)$ where $\Delta v$ is the difference of velocity between the two layers and $\lambda$ the wavelength of the perturbation. In practice, numerical simulations are limited by diffusivity and the minimum resolvable structure, inevitably stunting the instability at small $\lambda$. At the other end of the scale, the development of instabilities with large wavelengths can be hampered by their advection in the flow. The dynamical timescale  can be estimated by $\tau_{dyn} \sim a/c_s$ where $c_s$ is the post-shock sound speed, which is of the order of the wind velocity $v_\infty$ in a strong adiabatic shock. Hence, the scale of the perturbations may be expected to be limited to $\lambda/a <  \Delta v /v$. For two identical winds with terminal velocities of 2000 km\, s$^{-1}$ and $a=1$ AU, $\tau_{dyn}\simeq 6.8\times 10^4 $s$=2.2\times 10^{-3}$ yr. 

We performed a set of simulations with $\eta=1$, increasing the velocity $v_{\infty 1}$ of the first wind to investigate the impact of the KHI in the adiabatic case. The mass loss rate $\dot{M}_1$ was simultaneously decreased and the Mach number ${\cal M}_1$ of the wind was kept equal to 30. The size of the domain is $8a$ and the resolution is $n_x=128$ with 5 levels of refinement. The simulations were run up to $t=600 \tau_{dyn}$. A steady state is reached well before the end of the simulation, as determined by looking at the time evolution of the total r.m.s. of the density or velocity perturbations over the whole simulation domain. Restricting ourselves to this steady state interval, which we checked to be much longer than the advection time along the contact discontinuity, we then computed the time average of the velocity r.m.s. for each cell of the domain. We used the median value over the same time period as our reference. The purpose was to quantify the saturation amplitude of the perturbations. 

\begin{figure*}
\centering
\includegraphics[width = .29 \textwidth]{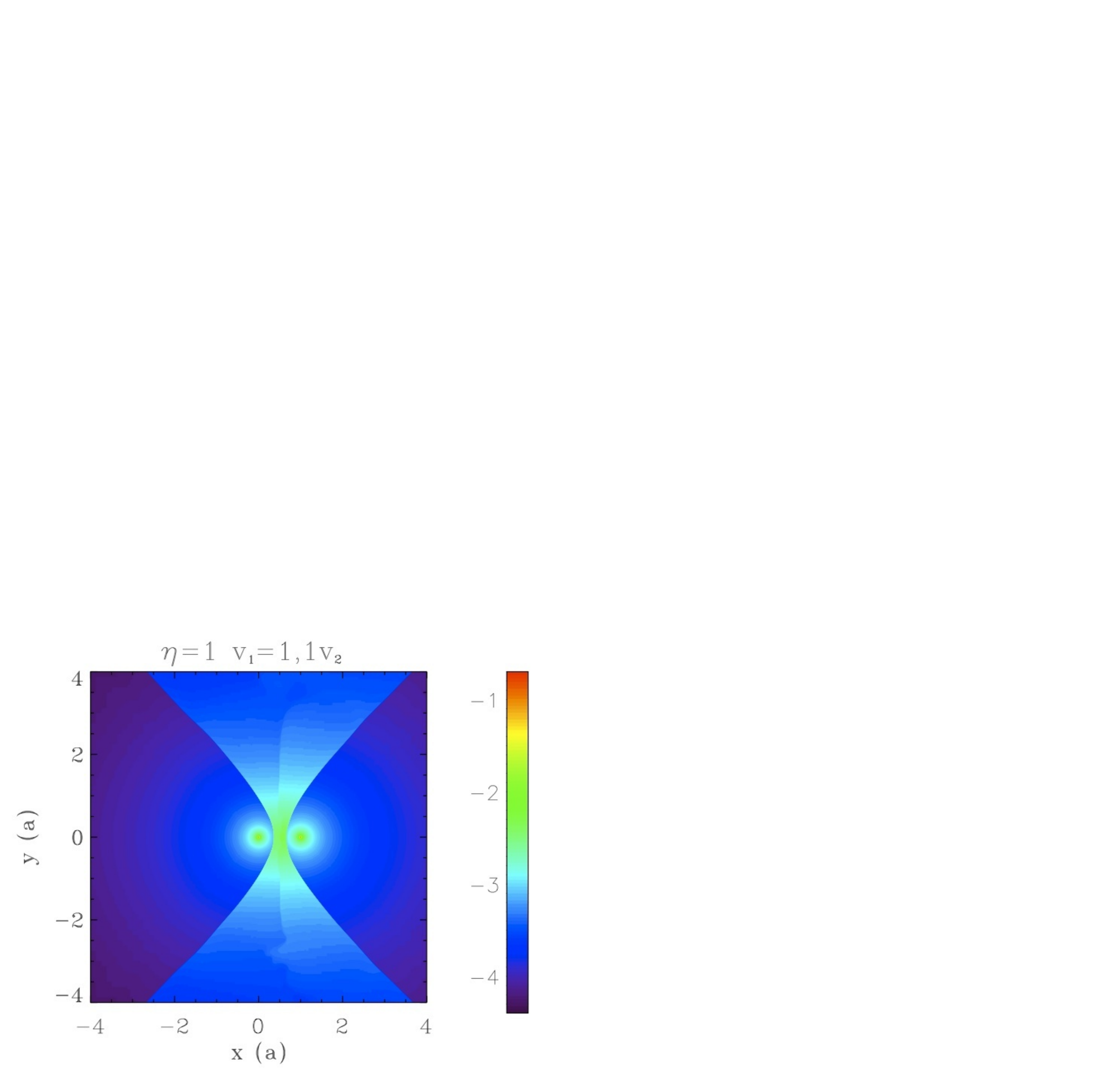}
\includegraphics[width = .29\textwidth]{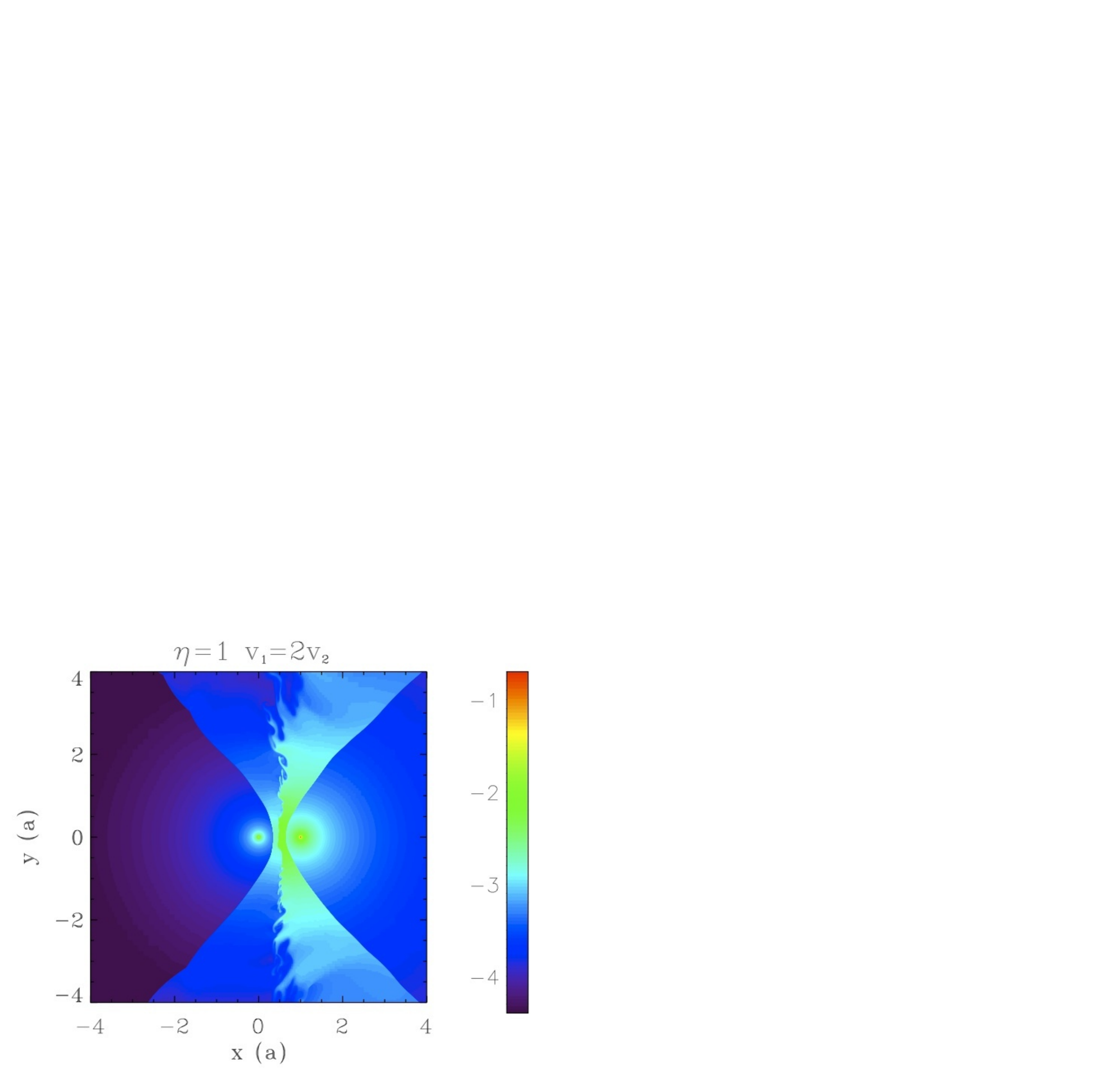}
\includegraphics[width = .29 \textwidth]{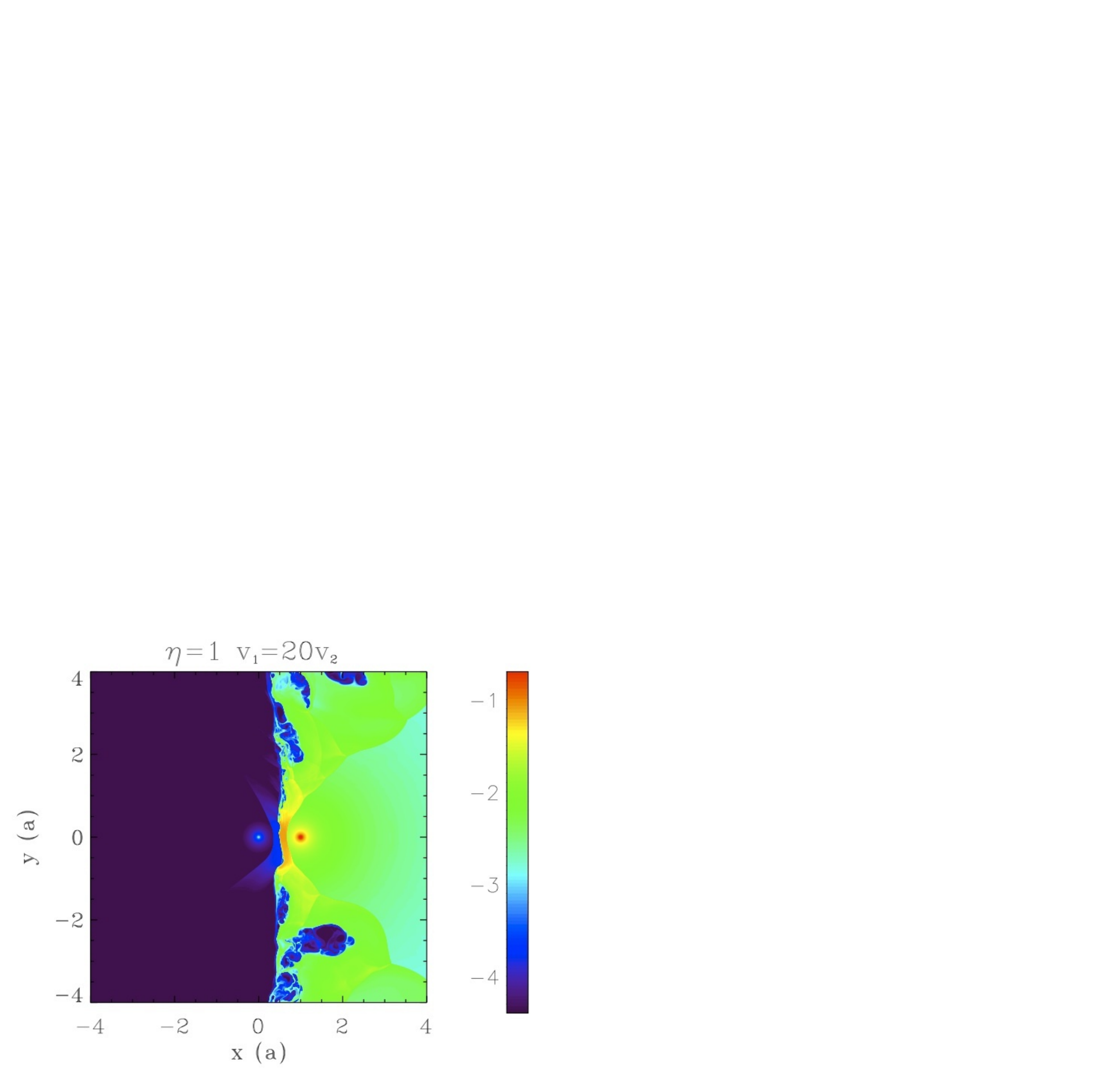}
\includegraphics[width = .29\textwidth]{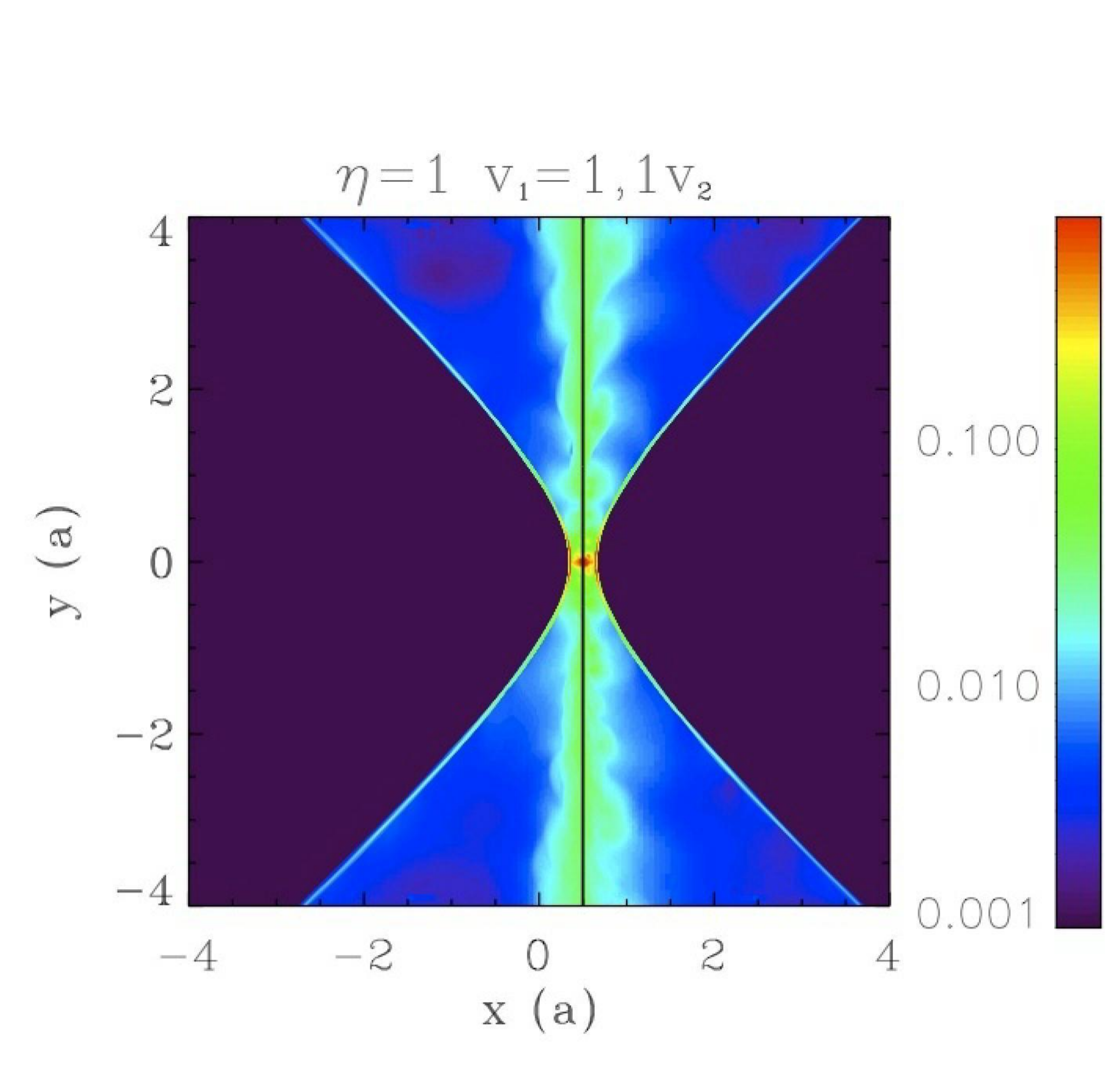}
\includegraphics[width = .29\textwidth]{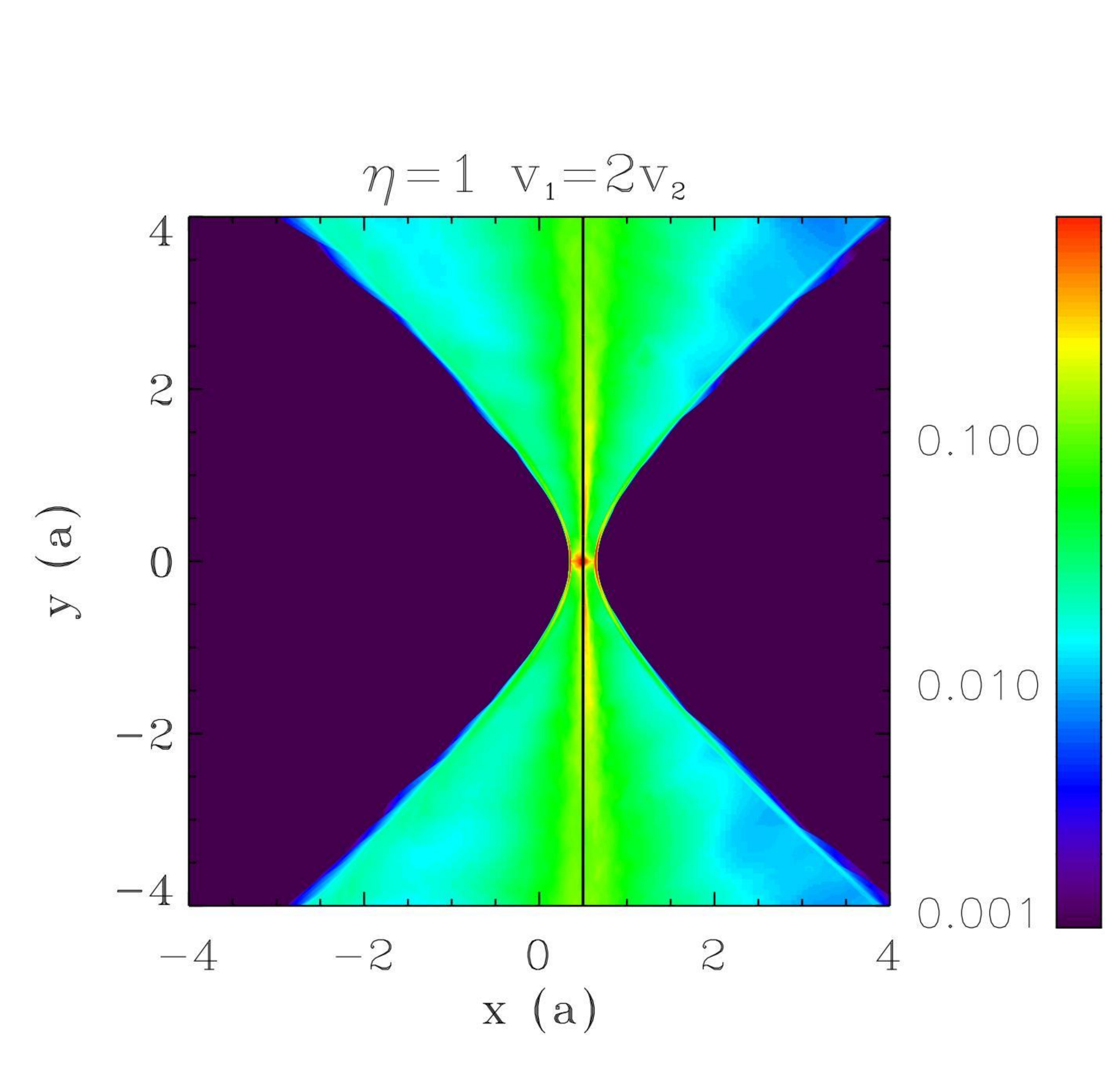}
\includegraphics[width = .29\textwidth]{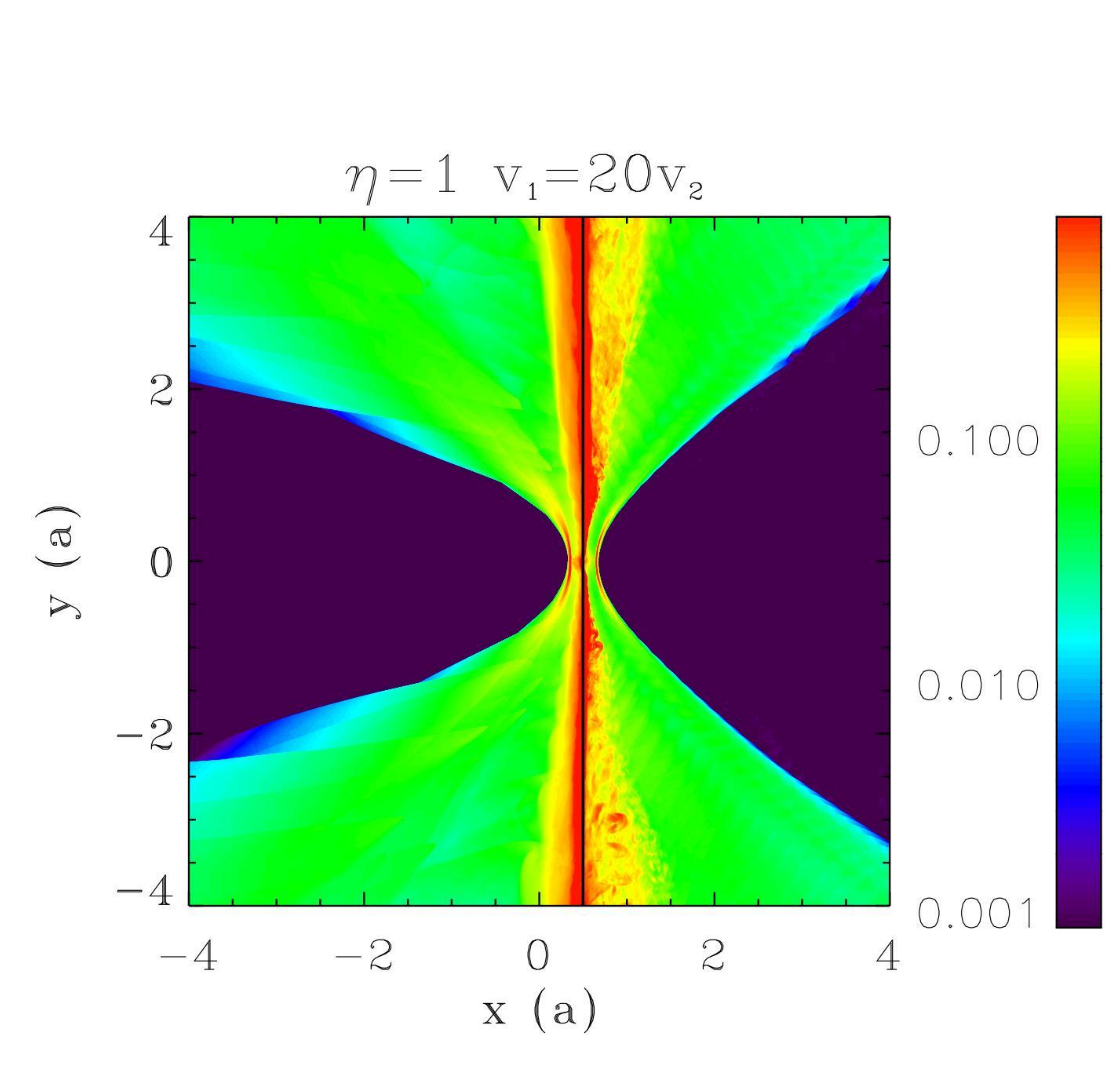}
\caption{Development of Kelvin-Helmholtz instability in the adiabatic case for $\eta=1$. Upper panel: density maps from left to right : 
$v_{\infty 1}=1.1v_{\infty 2},\rho_1=0.91\rho_2$; $v_{\infty 1}=2v_{\infty 2},\rho_1=0.5\rho_2$;  $v_{\infty 1}=20v_{\infty 2},\rho_1=0.05\rho_2$. Lower panel : r.m.s. of the velocity perturbations on a logarithmic scale. The fastest wind originates from the star on the left hand side.}
\label{fig:KH_eta1} 
\end{figure*}

The results are shown in  Fig.~\ref{fig:KH_eta1}. The upper panels gives the density maps for the different cases while the corresponding lower panels show the time average of the r.m.s of the velocity fluctuations. No instabilities are present when the two winds are exactly identical, as expected since there is no velocity shear. Introducing a 10$\%$ difference in the velocity of the winds leads to low amplitude perturbations that are significant only close to the contact discontinuity. A dominant wavelength can be identified, probably because growth for such a weak velocity shear is restricted to a small domain by diffusivity at short wavelengths and advection at long wavelengths. The r.m.s. of the velocity and density perturbations saturates at about 10$\%$. When $v_{\infty 1}=2v_{\infty 2}$ small scale eddies are visible. They are stretched in the direction of the flow. The position of the shocks is barely affected by the instability. The perturbations affect a larger zone on both sides of the contact discontinuity but their amplitude remains around a few tens of percent r.m.s., somewhat higher for the density than for the velocity perturbations.  When $v_{\infty 1}=20v_{\infty 2}$ (fourth panel) the instability has become non-linear judging by the 100$\%$ r.m.s. of the velocity (and density) fluctuations. The location of the contact discontinuity fluctuates significantly yet the region with the strongest r.m.s. is not much wider than for the previous cases. We also investigated in this last case whether keeping the wind temperature constant as $v_{\infty 1}$ is varied, instead of keeping ${\cal M}_1$ constant, led to differences. The outcome was similar.

\begin{figure*}
\centering
\includegraphics[width = .29 \textwidth]{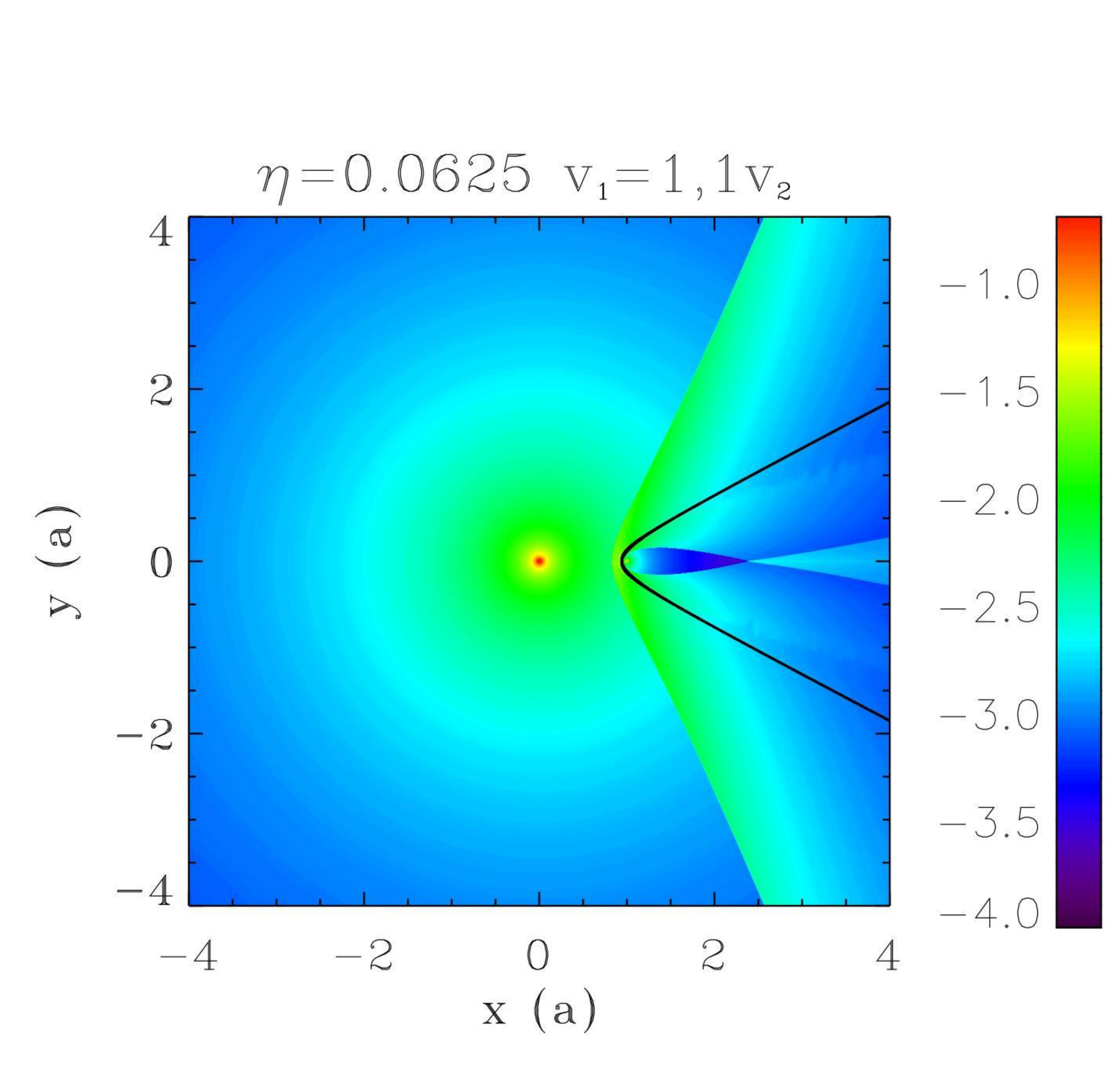}
\includegraphics[width = .29\textwidth]{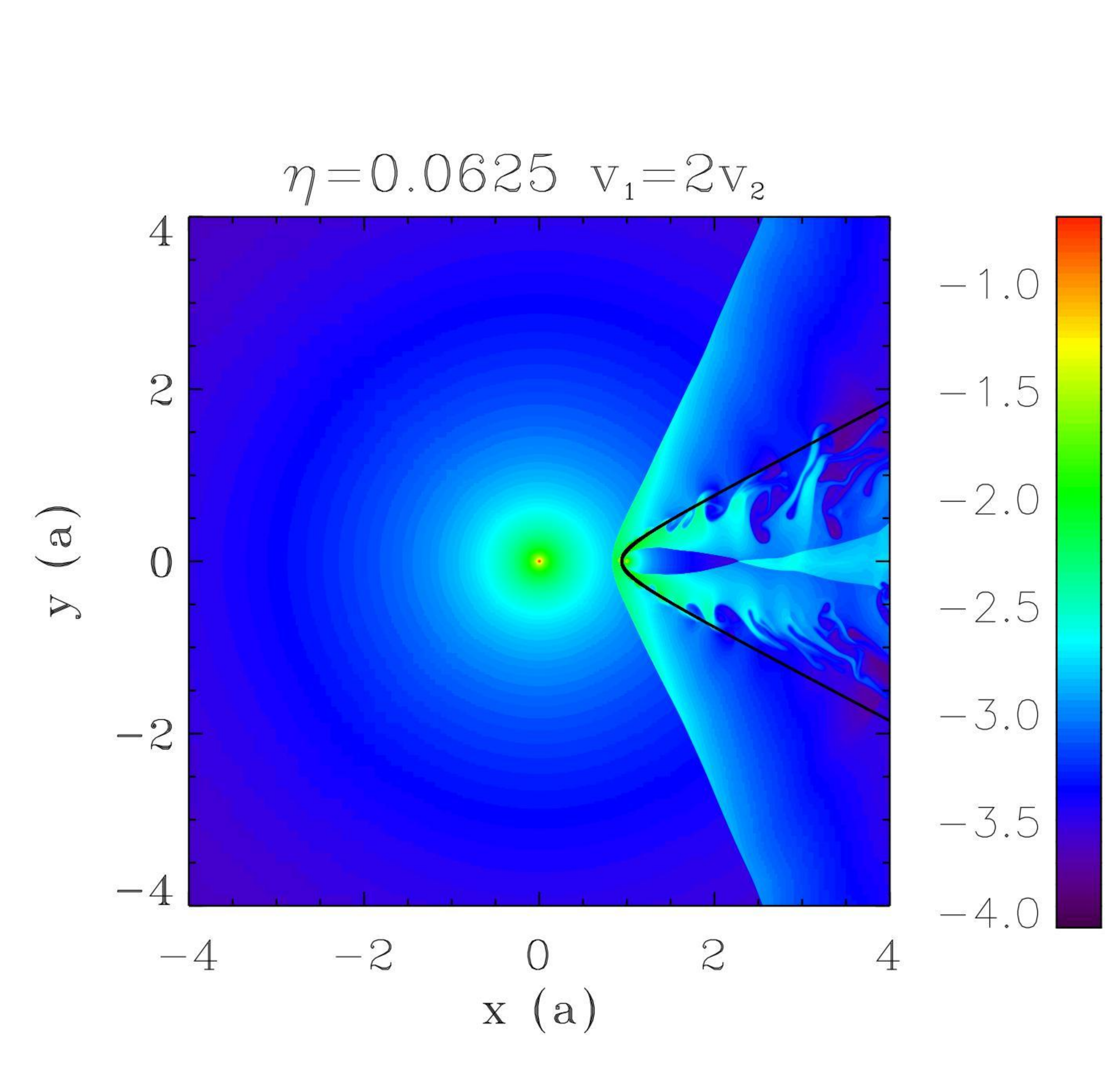}
\includegraphics[width = .29 \textwidth]{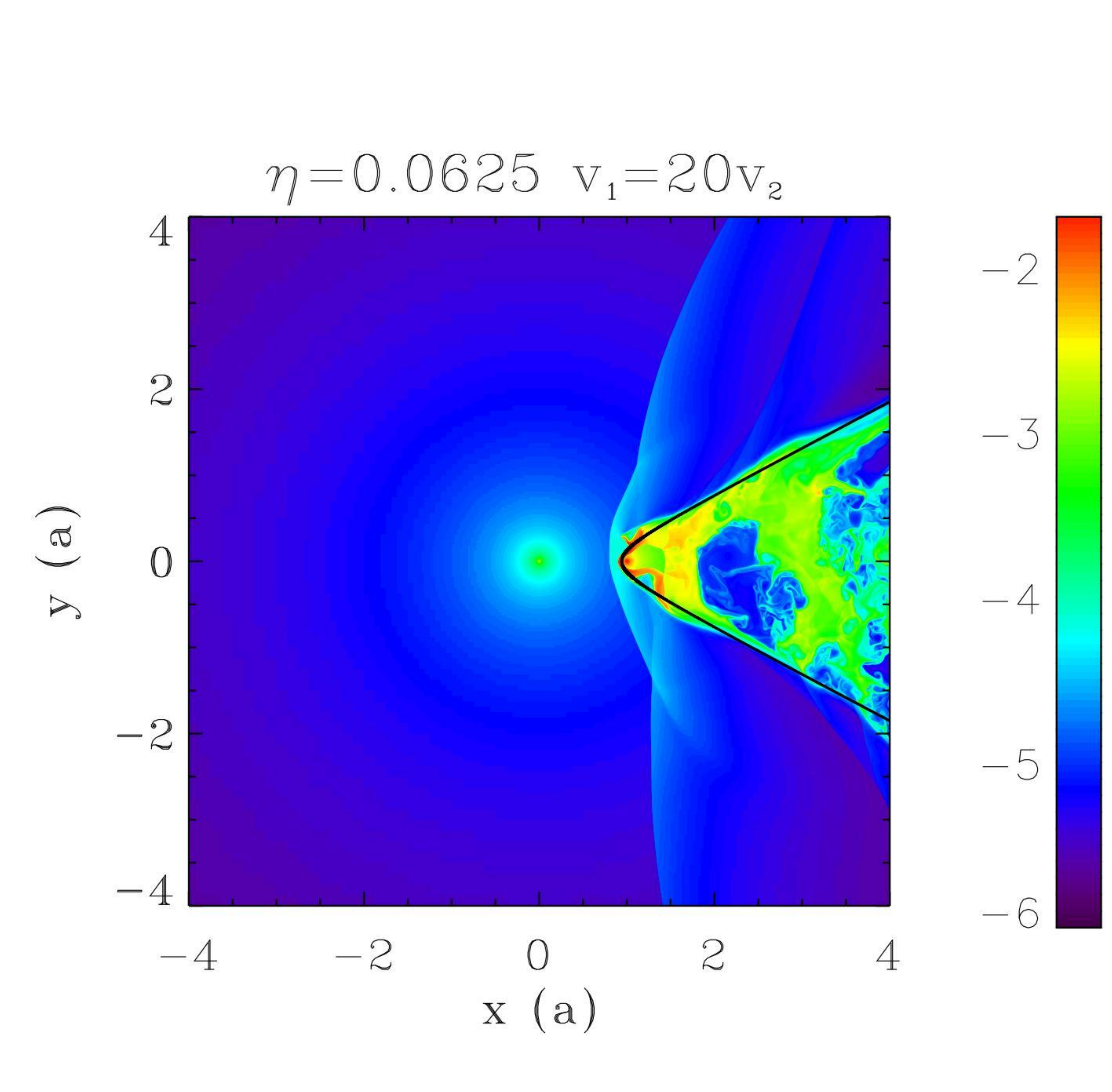}
\includegraphics[width = .29\textwidth]{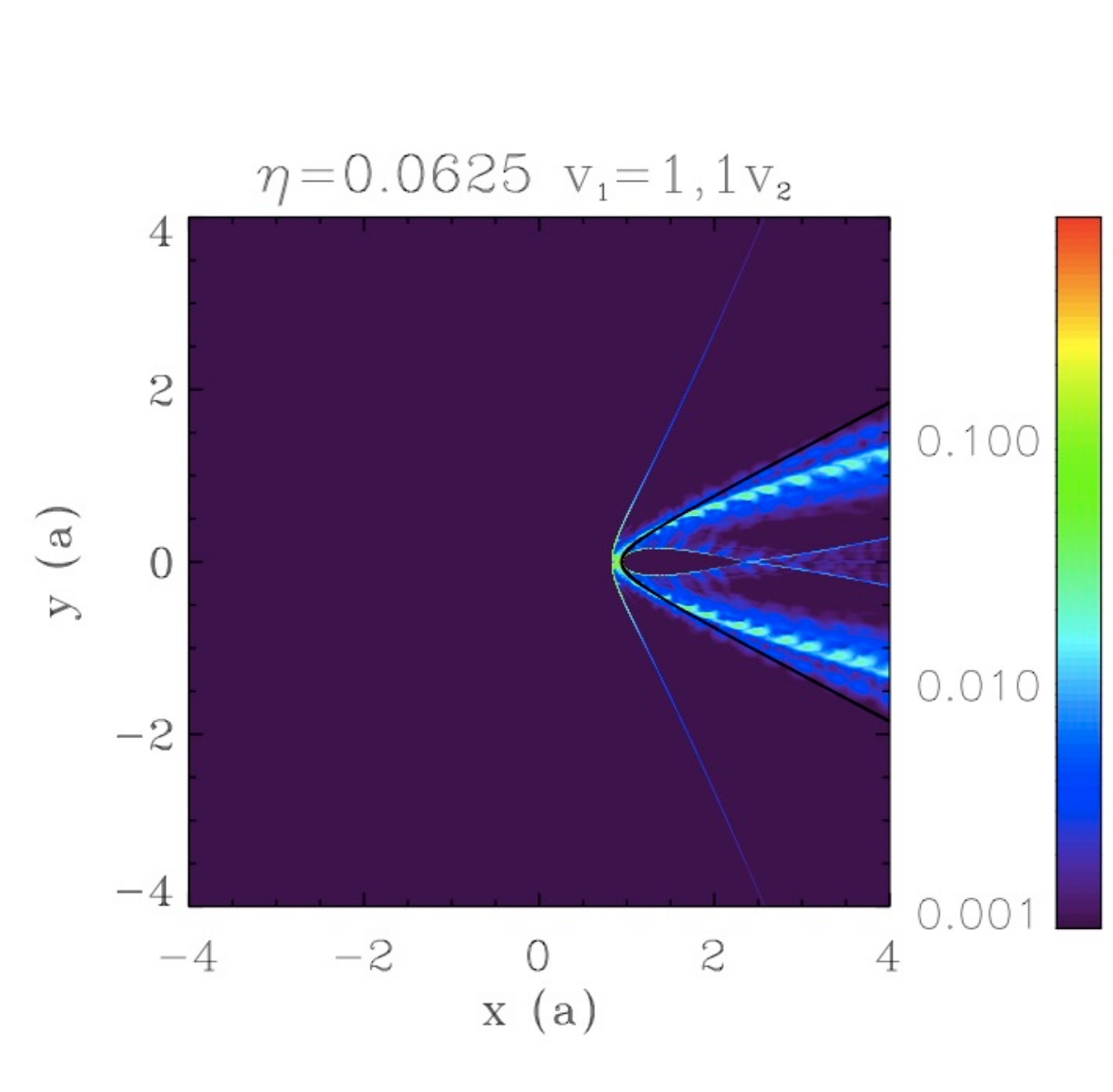}
\includegraphics[width = .29\textwidth]{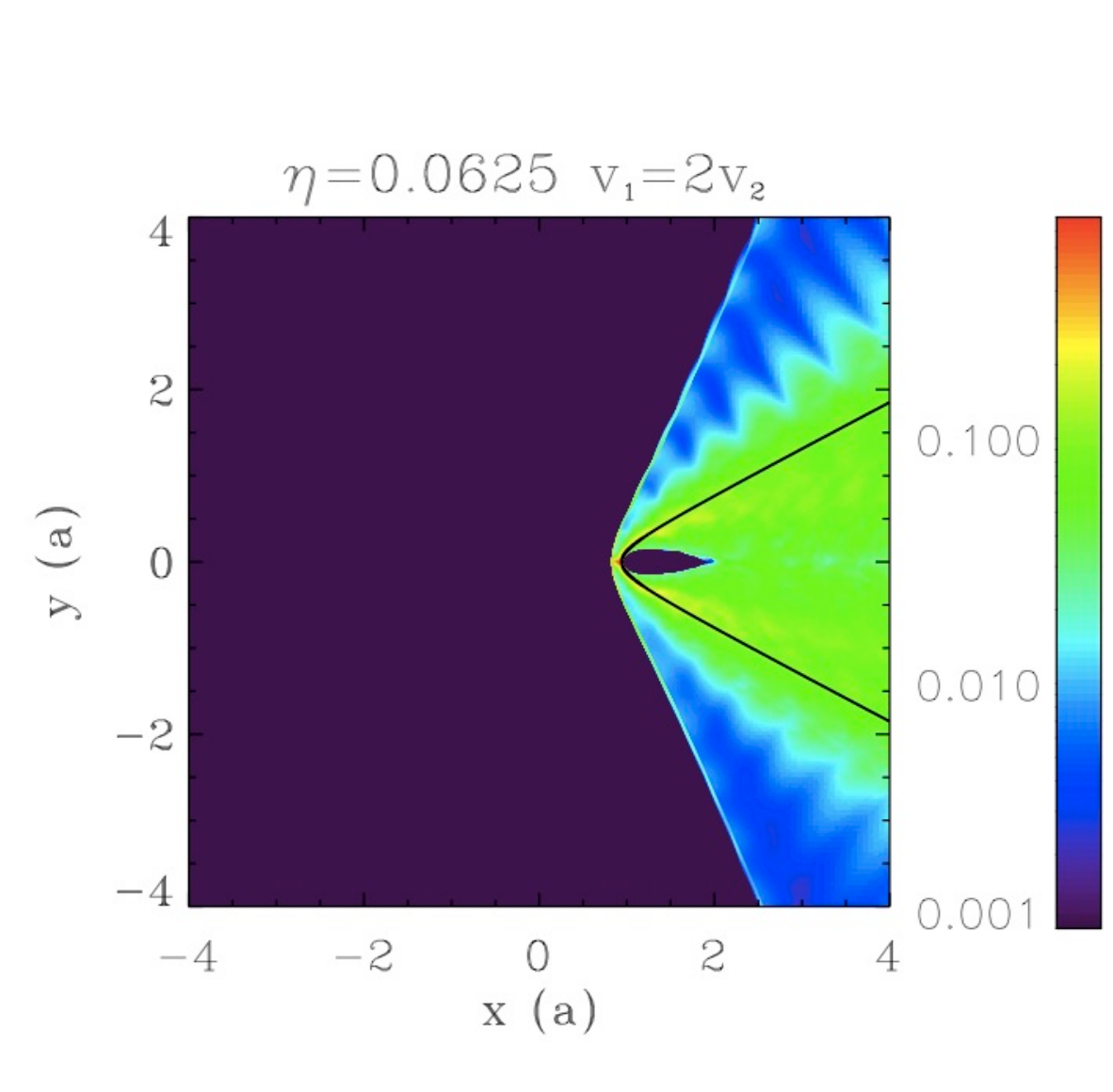}
\includegraphics[width = .29\textwidth]{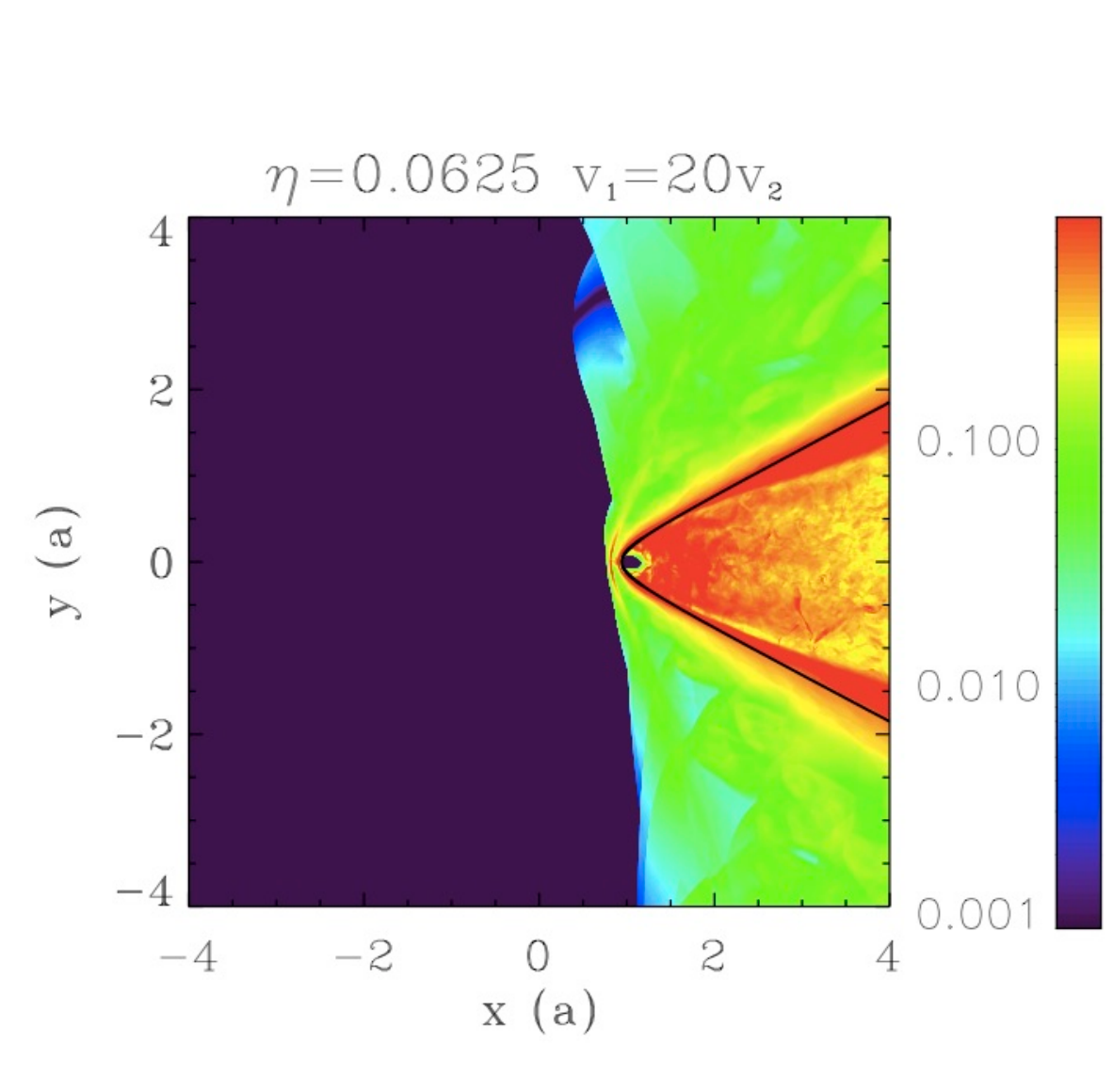}
\caption{Same as Fig.~\ref{fig:KH_eta1} but for $\eta=1/16=0.0625$.}
\label{fig:KH_eta16} 
\end{figure*}

A similar set of simulations was performed with $\eta=1/16=0.0625$ (Fig~\ref{fig:KH_eta16}). There is no velocity shear or contact discontinuity when $v_{\infty 1}=v_{\infty 2}$, even in the case $\eta\neq 1$. This can be proven as follows. The Bernouilli constant (Eq.~\ref{eq:Bernouilli}) has the same value in both shocked region when $v_{\infty 1}=v_{\infty 2}$, so the densities are identical at the contact discontinuity (where pressures equalise) on the line-of-centres. The gas is polytropic with $P\equiv K \rho^{-\gamma}$ and $K$ constant in each region.  Writing that $\rho$ and $P$ are equal on both sides of the contact discontinuity on the line-of-centres requires that $K$ has the same value in both shocked regions. Therefore, $\rho_{1s}=\rho_{2s}$ along the contact discontinuity. Using that the Bernouilli constant is the same in both shocked regions then proves that $v_{1s}=v_{2s}$ at the contact discontinuity. Actually, there is no discontinuity in this case. The simulation with $v_{\infty 1}=v_{\infty 2}$ confirms that there is no velocity shear and that the KHI does not develop. When $v_{\infty 1}=1.1v_{\infty 2}$ only weak perturbations are seen, limited to a small region close to the contact discontinuity. A dominant wavelength can be identified as in the case $\eta=1$. When $v_{\infty 1}=2v_{\infty 2}$ the center line of the perturbations approximately matches the shape of the unperturbed contact discontinuity. The first shock is not affected by the instability. The velocity perturbations affect all the region of the shocked second wind and part of the shocked wind of the first star. The density perturbations have a higher r.m.s. than the velocity perturbations, reaching close to 100$\%$ close to the contact discontinuity. The velocity perturbation are strong when $v_{\infty 1}=20v_{\infty 2}$ and are mostly confined to the shocked second wind. High r.m.s. density fluctuations extend to the first wind, distorting slightly the first shock. (The sawtooth appearance of the wings in the $v_{1\infty}=v_{20\infty}$ r.m.s. maps are an artefact of the limited time range over which the average was done.) The backward reconfinement of the wind of the second star is affected by the instability, occurring much closer to the second star than in the case with equal wind velocities.

The KHI modifies the interaction region as soon as the wind velocities are slightly different. The simulations suggest that the relative amplitude of the perturbations becomes significant when $v_1\ga 2 v_2$, although we cannot rule out that limited numerical resolution does not impact the growth of the instability for smaller velocity shears. The instability does not erase completely the contact discontinuity. However, the turbulent motions tend to smooth out the initial structures in the region of the wind with the smaller velocity.

\subsection{Isothermal equation of state: thin shell instabilities\label{isothermal}}
When thermal support in the shocked zone is too weak, the shell becomes thin and unstable. This occurs for instance when the adiabatic index is decreased \citep{1993ApJ...407..207M}. More realistic numerical simulations including radiative cooling functions also show the shocks become thinner and unstable as cooling increases (\citealt{Stevens:1992on,2009MNRAS.396.1743P}, but see \citealt{1998MNRAS.298.1021M}). The instability is usually referred to as `thin shell instability' although several physical mechanisms may be at work, including the KHI. The non-linear thin shell instability (NTSI, \citealt{1994ApJ...428..186V}) is found in hydrodynamical simulations when the thin shell is moved away from its rest positions by perturbations with an amplitude at least greater than the shell width \citep{1996NewA....1..235B}. The instability is due to an imbalance in the momentum flux within the shell as shocked fluid moves towards opposing kinks. The transverse acceleration instability (TAI, \citealt{1993A&A...267..155D,1996ApJ...461..927D}) occurs when at least one of the colliding flows is divergent and assumes an infinitely thin shell. Both linearly unstable breathing and bending modes are found. The breathing mode is due to the acceleration of the flow along the shell whereas the bending mode arises from the mismatch in ram pressure of the wind impacting each side of the thin shell when it is displaced from its equilibrium value.

We studied the growth of thin shell instabilities in colliding wind binaries using 2D simulations with an isothermal equation of state. Initial investigations showed that the thin shock structure (\S\ref{2dstudy}) becomes unstable only if there are a sufficient number of cells available ($\ga 4$) to resolve the shock structure. The minimum number of cells required is even larger if a highly diffusive solver is used. Low resolution simulations without mesh refinement  ($256\times 256$ cells) do not resolve the shock structure and stay stable. We decided to use those steady state solutions as the initial input for simulations at higher resolution, so as to be able to study in as much as possible the initial linear growth phase of the instabilities. The winds are chosen to have identical velocities in order to exclude any seeding by the KHI (\S\ref{KH}).

\begin{figure}
\centering
\includegraphics[width = .5\textwidth]{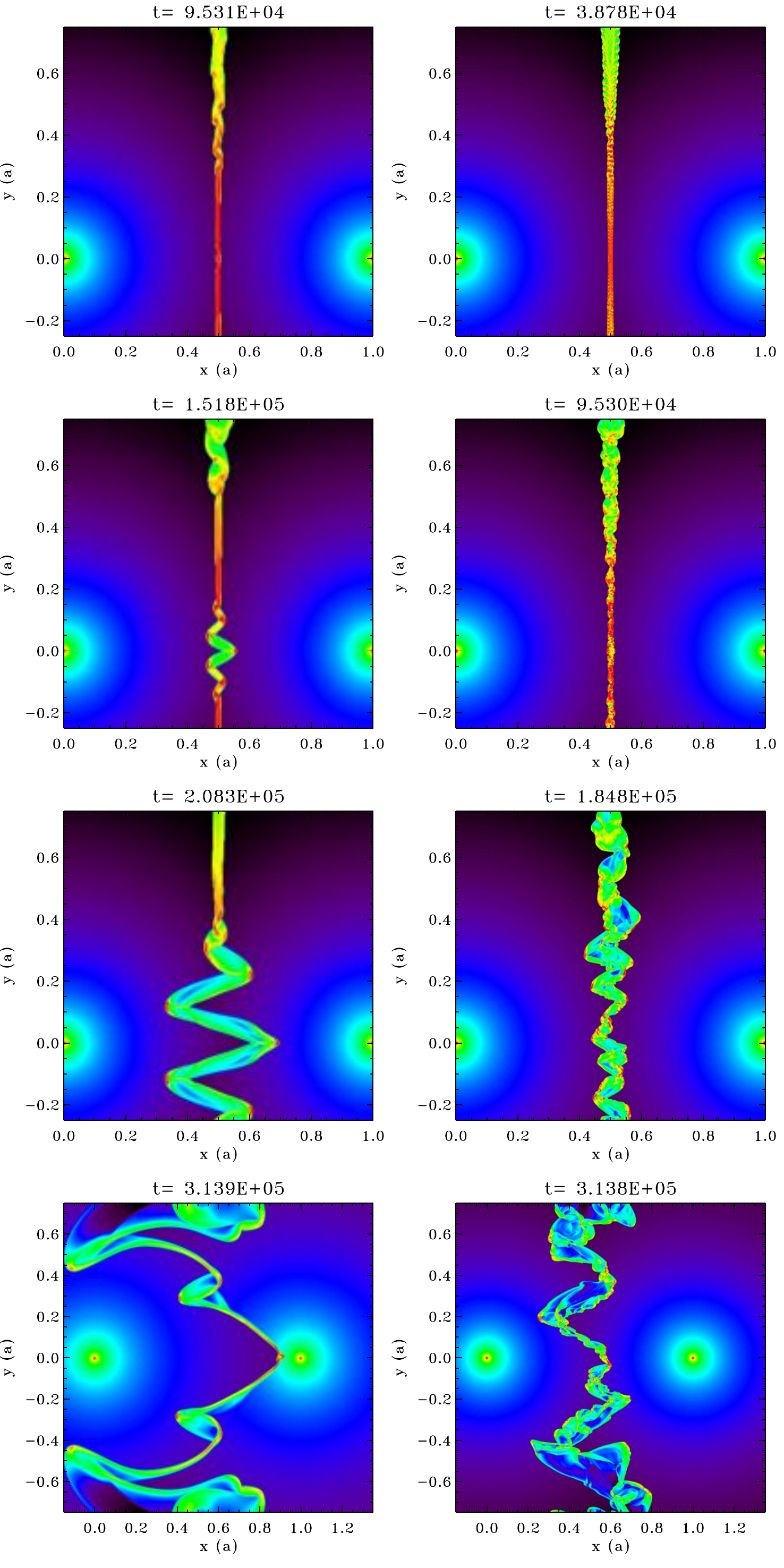}
\caption{Density maps showing the evolution of a 2D colliding wing binary when $\eta=1$ and $\gamma=1.01$. Time is given in seconds. $t=0$ corresponds to the restart  at high resolution of an initial low resolution simulation (256 cells, no mesh refinement). On the left panel there is one level of refinement (maximum resolution equivalent to 512 cells), on the right panel there are four levels (maximum resolution equivalent to 4096 cells).}
\label{fig:evolution} 
\end{figure}

The evolution of a colliding wind binary with $\eta=1$, identical velocities and an isothermal equation of state is shown in Fig.~\ref{fig:evolution}. The size of the domain is $3a$. The left panels show the case with one level of mesh refinement, the right panels show the case with four levels. At low resolution (left panels), perturbations become visible away from the line-of-centres early in the simulation ($t=9.5 \times 10^{4}$ s). These perturbations grow slowly as they are advected, thickening the layer. At $t=1.5\times 10^{5}$\ s  another instability develops close to the binary with a growth rate faster than the advection rate and a distinct morphology. In this case matter piles up in the convex parts of the shell, which move steadily away from the initial shock position without the oscillatory behaviour seen in the wings. At the end of the simulation ($t=3.1\times 10^{5}$ s) the colliding wind region is dominated by these large scale perturbations. At higher resolution (right panels), the initial instability appears earlier and is also present closer to the binary axis. At $t=9.5\times 10^{4}$ s there already is a superposition of modes and one cannot define a unique wavelength any more. At $t=1.8\times 10^{5}$ s oscillations are present even on the binary axis and the structure is not symmetric any more. The final density maps shows a thicker shell with small scale structures. The oscillations are smaller than for the low resolution simulation at this time. The evolution at subsequent times shows comparable amplitudes in the oscillations at high and low resolution.

Similar behaviour was described by \citet{1998NewA....3..571B} in their simulations of stellar wind bow shocks. We tentatively associate the small amplitude instability that develops first, away from the binary axis, with the TAI. This is a linear instability that can be seeded by the initial numerical noise. The large amplitude instability that develops later on the binary axis is likely to be the NTSI. We examine below the supporting evidence.

\subsubsection{Evidence for the Non-linear Thin Shell Instability (NTSI)\label{ntsi}}

\begin{figure}
\centering
\includegraphics[width = .295\textwidth]{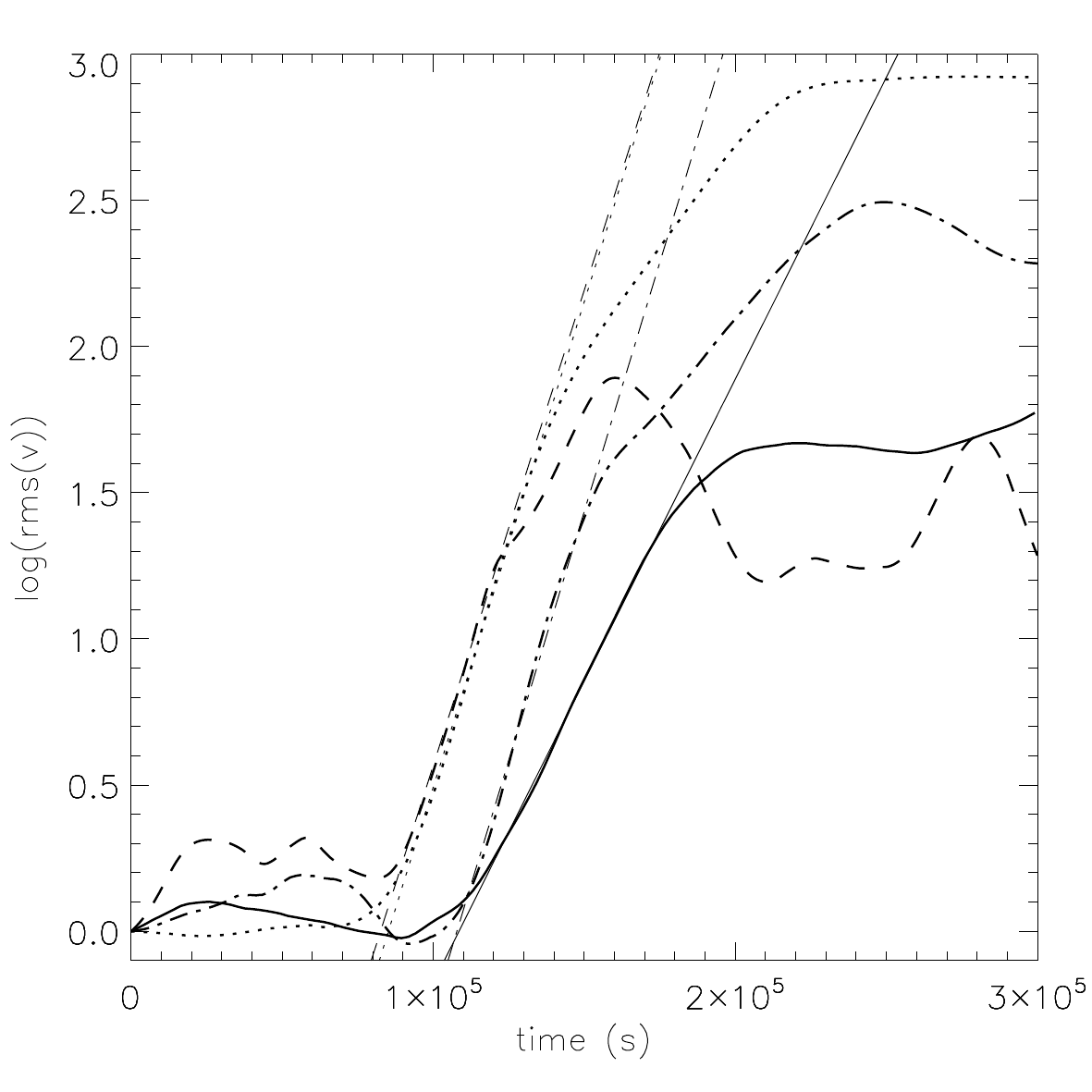}
\caption{Logarithm of the root mean square of the velocity on the line of centres as a function of time. The curves represent maximal resolutions of 512 (dotted), 1024 (dot-dashed), 2048 (dashed) and 4096 (solid) cells per dimension. The thin straight lines show the fits to the linear phase for each resolution.}
\label{fig:NTSI} 
\end{figure}

The NTSI shows the highest growth rate for perturbations of order of the shell width $L$. The  theoretical estimate is $\tau_{th}=L/c_s=2.0\times 10^{4}$ s~\citep{1994ApJ...428..186V} for the parameters appropriate to our simulations, smaller than the advection timescale ($\tau_{dyn}\simeq 6.8\times 10^4 $, increasing near the binary axis as the flow velocity in the shocked region goes to zero on axis). Hence, the fastest growing mode of the NTSI should be seen, independently of the numerical resolution, as long as the shell is resolved. We compared this estimate with the time evolution of the velocity perturbations in four simulations with 1, 2, 3 and 4 levels of refinement, using an exact Riemann solver. For each simulation, we computed the r.m.s. of the velocity for a line of cells along the binary axis, where the NTSI is presumed to dominate. We normalised the data to the value at the same arbitrary reference time taken close to the beginning of each simulation. The r.m.s. were smoothed to suppress small wavenumber perturbations that appear at high resolutions. The logarithm of the r.m.s is shown in Fig.~\ref{fig:NTSI}. The shell readjusts to the higher numerical resolution up to $t\simeq  9.5\times 10^{4}$\, s. Close inspection of the density maps reveals the presence of density fluctuations on the scale of the shock width during this transition. This numerical relaxation triggers the NTSI close to the binary axis (left panels of Fig.~\ref{fig:evolution}). In the simulations with highest numerical resolution (right panels of Fig.~\ref{fig:evolution}) the NTSI develops in regions that are already perturbed by the growth of the first instability (most likely the TAI, see \S\ref{tai}). These fast growing perturbations may contribute to trigger the NTSI. The NTSI moves the shock away from its rest position as the bending modes are amplified and mass collects at the extrema \citep{1994ApJ...428..186V}. The exponential growth timescale  estimated from fitting the r.m.s values are $\tau\approx 3.1\times 10^4$, 2.9$\times 10^{4}$, 4.5$\times 10^{4}$ and 4.7$\times 10^{4}$\, s for increasing resolutions (mesh refinement). There is an increase of 50\% of the measured growth  timescale whereas the cell size (and therefore the available wavelength range potentially accessible) increases by a factor 16. This is in reasonable agreement with the theoretical value and the expected behaviour with changing resolution, confirming that the NTSI is triggered in our simulations.  Fig.~\ref{fig:NTSI} also shows that the saturation amplitude is somewhat smaller as the resolution is increased  (compare also the bottom left and right panels of Fig.~\ref{fig:evolution}) and that it converges  to a resolution-independent value.

\subsubsection{Evidence for the Transverse Acceleration Instability (TAI) \label{tai}}
The numerical simulations  show that the initial perturbations are preferentially located off the binary axis, have an oscillatory behaviour with a small wavelength and grow faster when the spatial resolution is increased (Fig.~\ref{fig:evolution}). The rapid development of these perturbations is consistent with a linear instability. These properties are reminiscent of the TAI. The TAI studied by  \citet{1993A&A...267..155D,1996ApJ...461..927D} is an overstability with an oscillation frequency of the velocity perturbations $\propto 1/\lambda$. The growth timescale is $\propto \sqrt{\lambda}$ and indeed smaller wavelength perturbations grow faster at higher resolution. \citet{1994ApJ...428..186V} noted that the growth is limited by pressure effects and that the TAI grows faster than the NTSI when
\begin{equation}\label{eq:TAIvsNTSI}
\frac{l}{R_s}> \frac{2\pi}{\mathcal{M}^2}\frac{R_s}{\lambda}
\end{equation}
Here, $l$ is the minimum distance along the contact discontinuity  ($l=0$ on the binary axis) beyond which the TAI can develop for a given wavelength $\lambda$. The relevant wavelengths are smaller than $R_s$ and larger than the shell width $L\sim R_s/\mathcal{M}^2$, with the smaller scales growing faster. The instability develops preferentially along the wings \citep{1998NewA....3..571B}. The presence of the TAI closer to the binary axis at the highest resolution may explain why the growth rate of the NTSI (see Fig. \ref{fig:NTSI}) does not perfectly match the theoretical value.

Despite the similarities, we could not formally identify the TAI. One difficulty is that we were not able to quantify the growth rates as several modes interact quickly and make the linear phase very short. Another is that we found that our initial velocity profile along the shock is inconsistent with the equilibrium solution proposed by  \citet{1993A&A...267..155D}. This was corrected by \citet{1998MNRAS.298.1021M} but they concluded that the set of equations used by \citet{1993A&A...267..155D} led to inconsistencies in the dispersion relations, casting doubt on the theoretical rates to expect. We suggest that it is not possible to neglect, as was done, the derivatives $\partial/\partial \theta$ in the equations ($\theta$ corresponds to the polar angle to the binary axis with the origin at the stagnation point), since there is a significant change in the azimuthal speed of the incoming flow as it is  decelerated and redirected along the shock. Although our results still support the presence in the simulations of some form of the TAI, the simulations also show that the saturation amplitude of this instability is low compared to the NTSI. In all the simulations we performed, the non-linear evolution was dominated by the large scale, high amplitude perturbations induced by the NTSI. At best, the TAI may play a role in the early stages as a seed instability for the NTSI, as described in \S\ref{ntsi}.

\subsubsection{Evolution with an initial velocity shear and at low $\eta$}\label{non_linear}
\begin{figure}
  \centering
  \includegraphics[width = .23\textwidth]{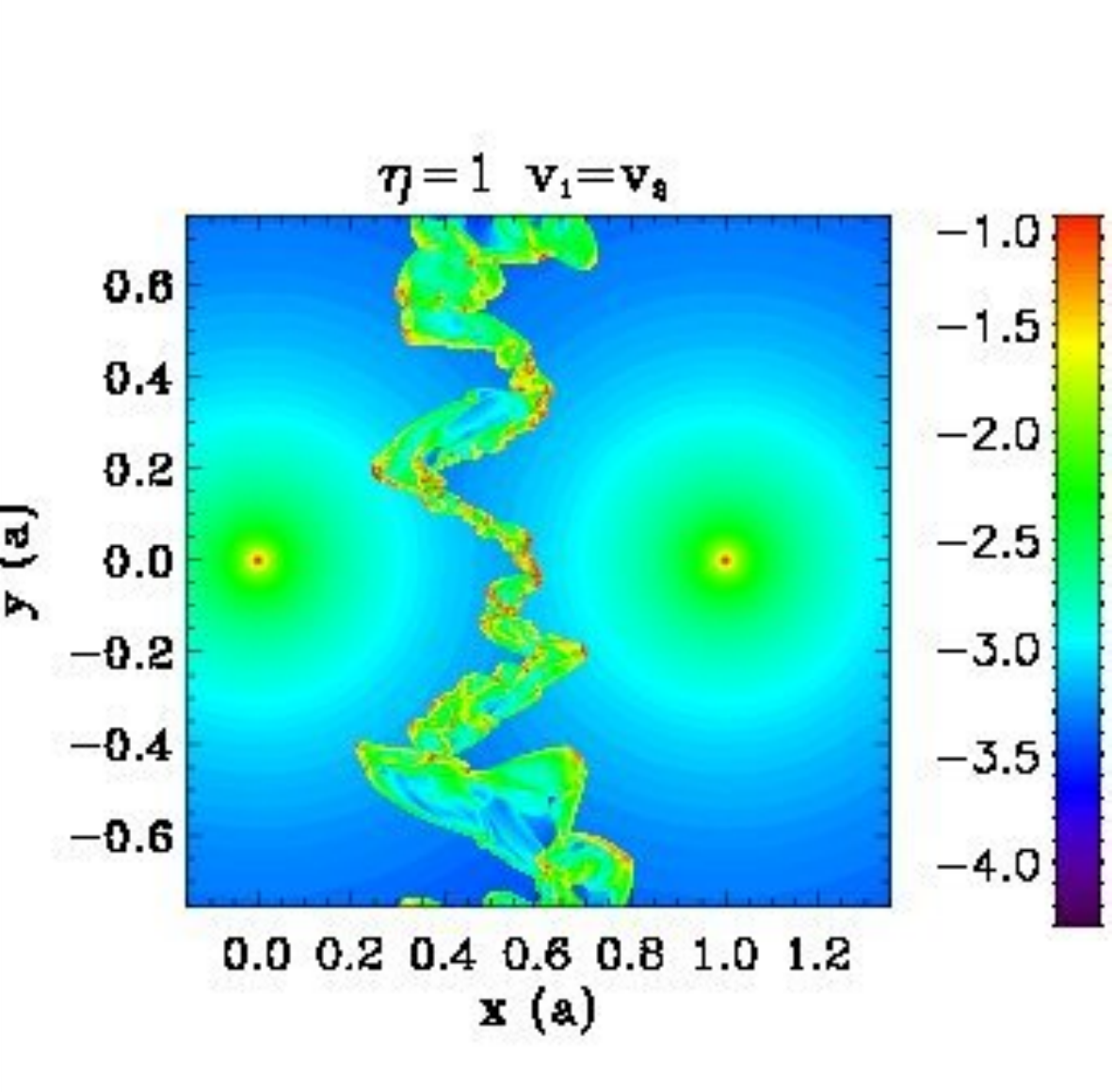}
  \includegraphics[width = .23\textwidth]{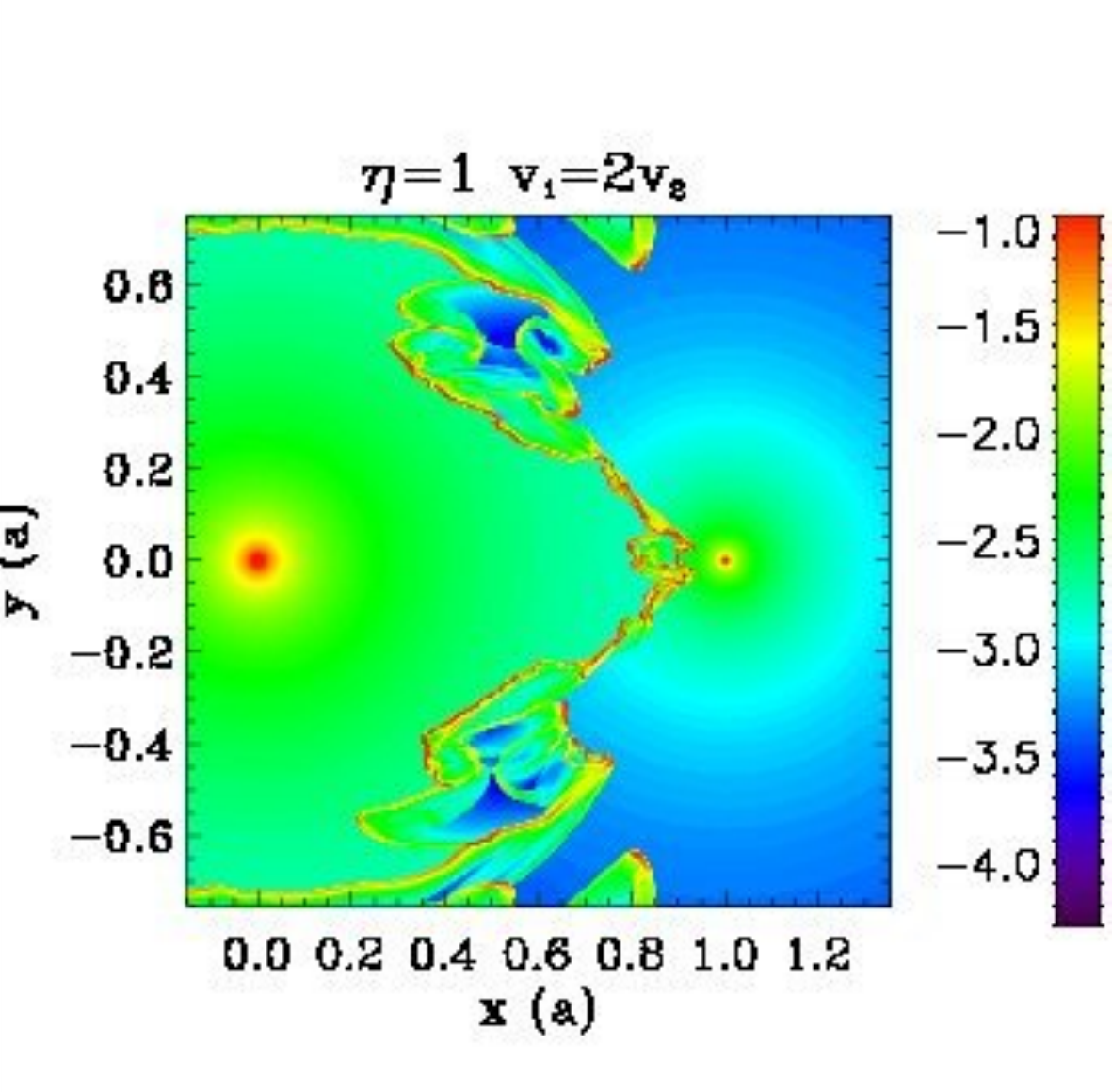}
  \includegraphics[width = .23\textwidth]{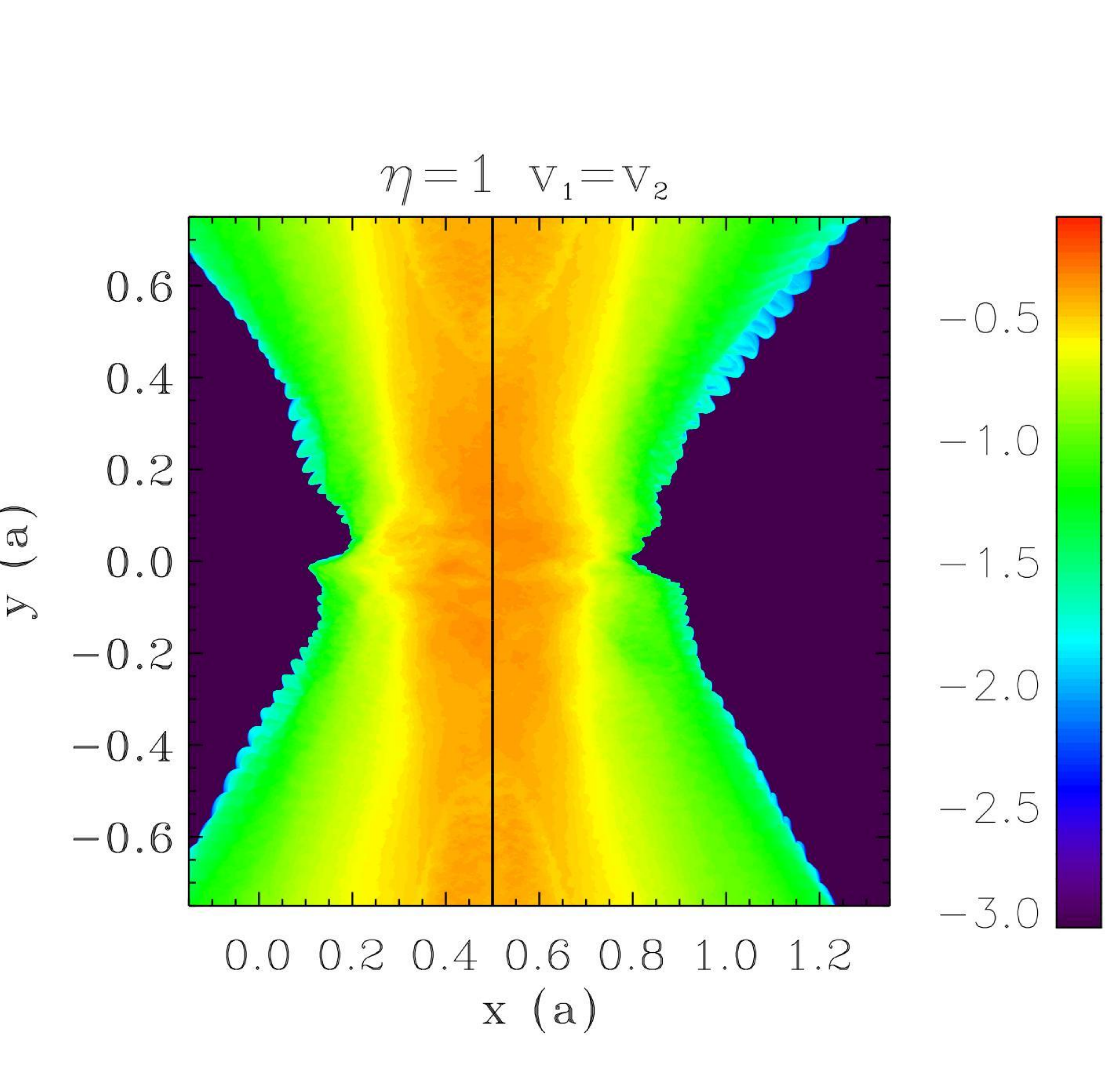}
  \includegraphics[width = .23\textwidth]{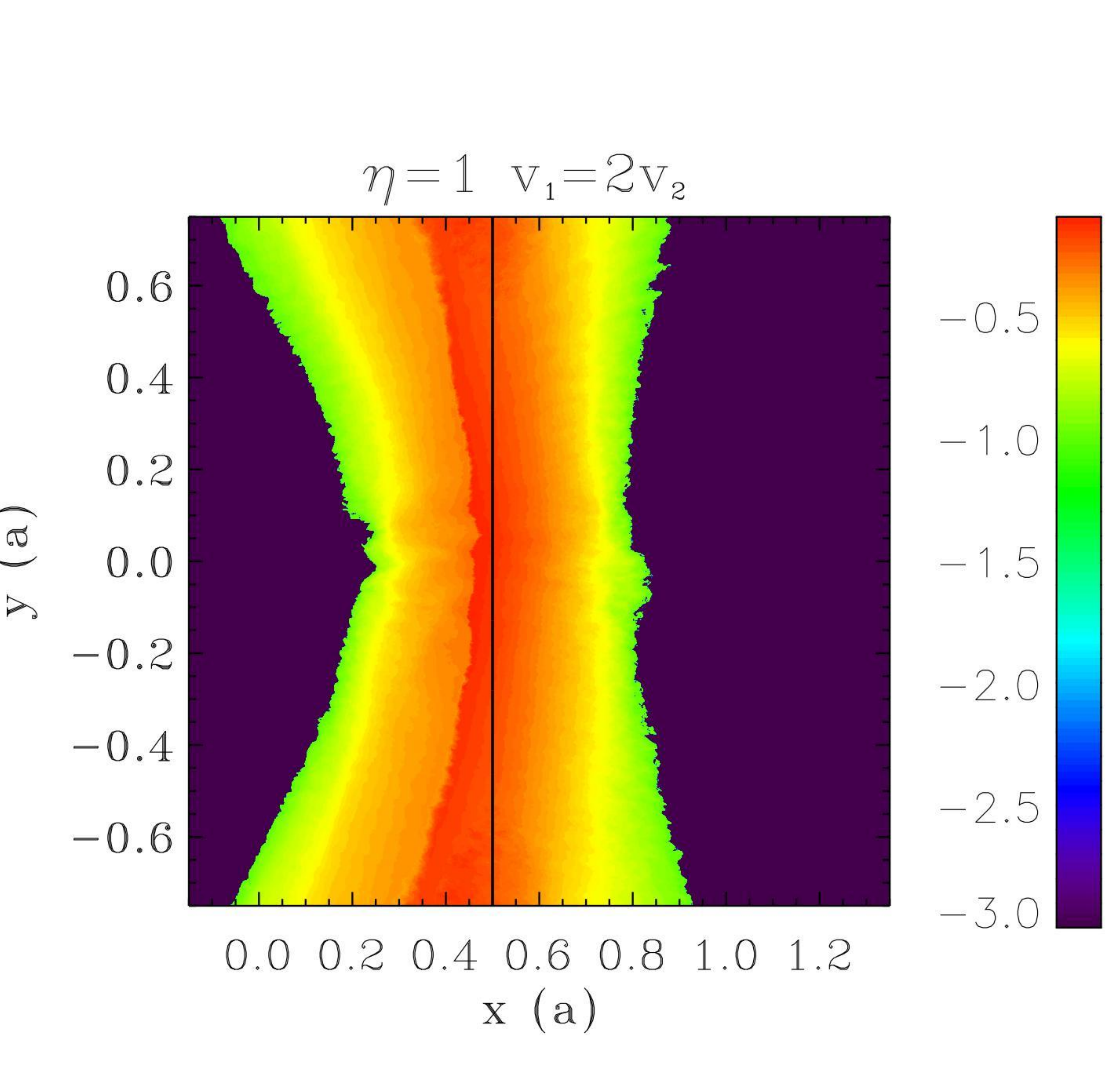}
  \includegraphics[width = .23\textwidth]{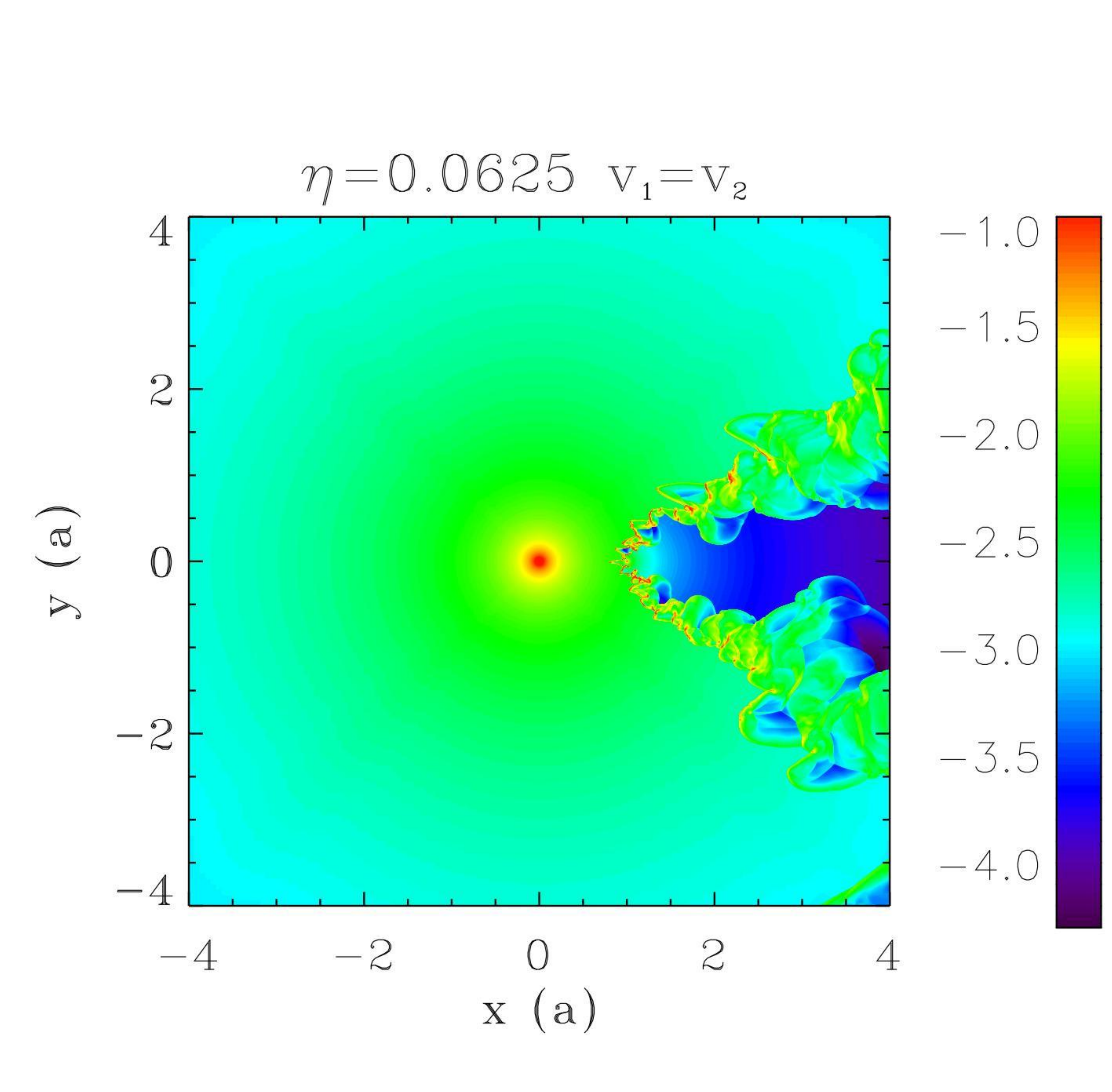}
  \includegraphics[width = .23\textwidth]{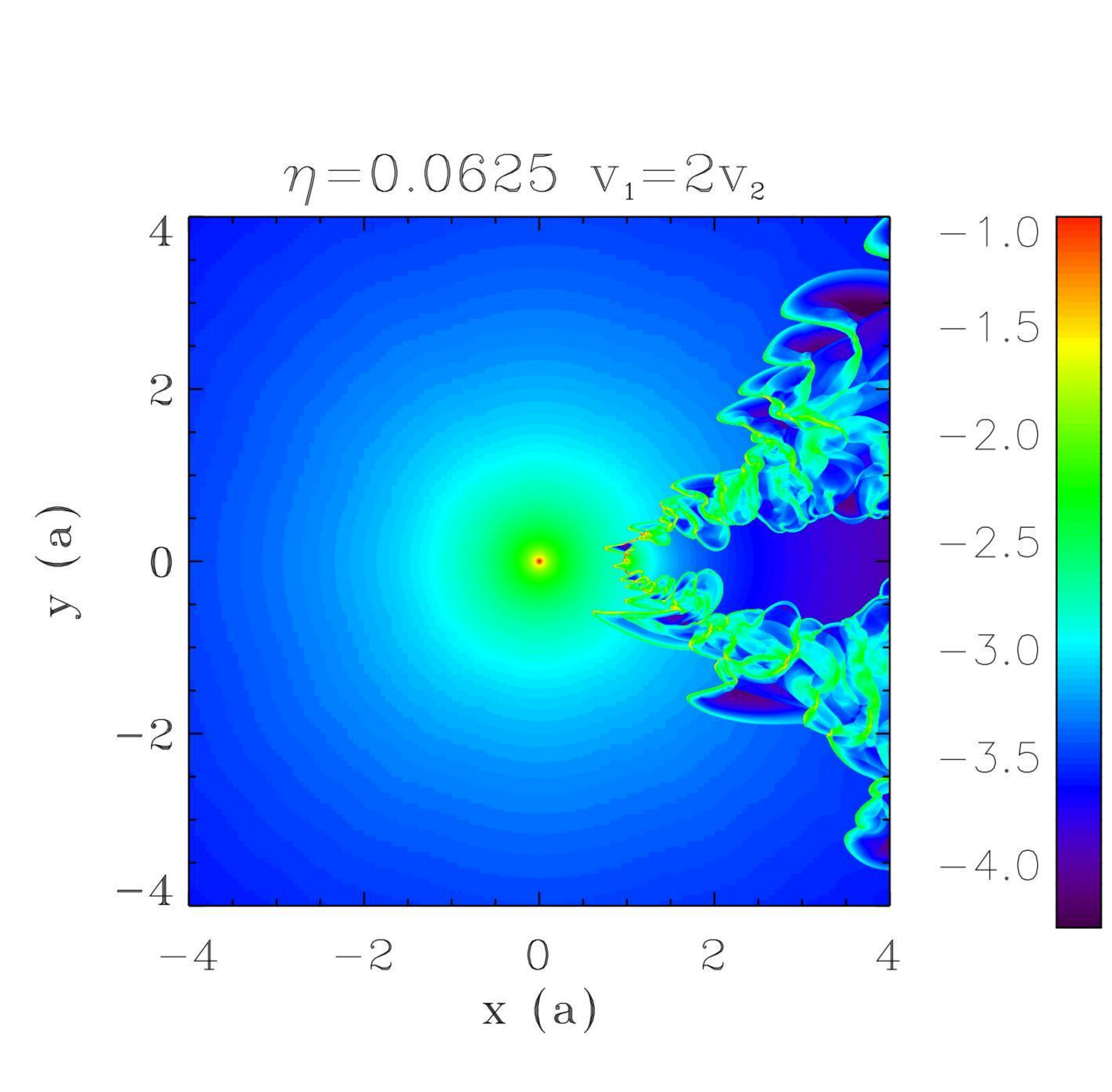}
  \includegraphics[width = .23\textwidth]{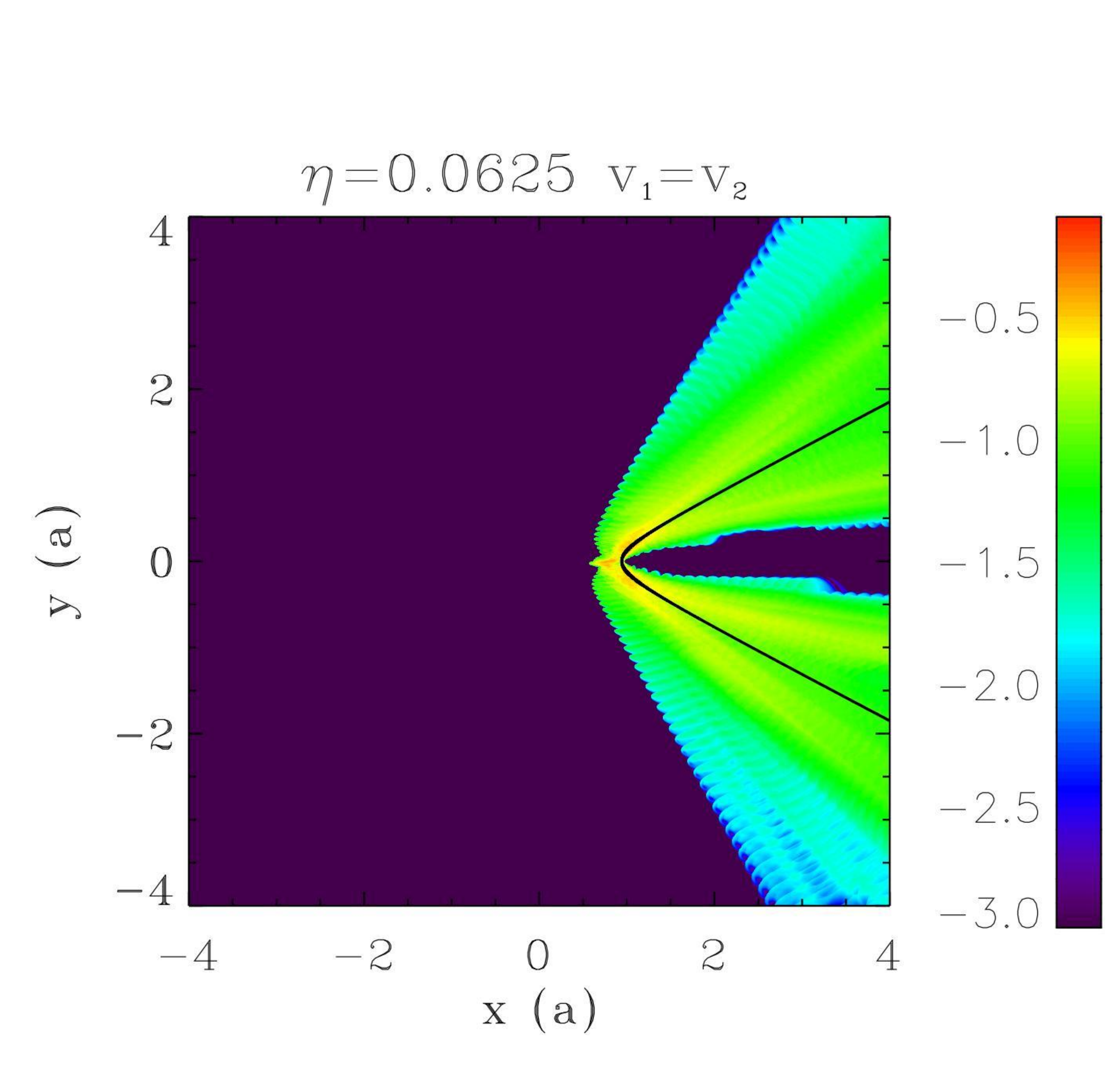}
  \includegraphics[width = .23\textwidth]{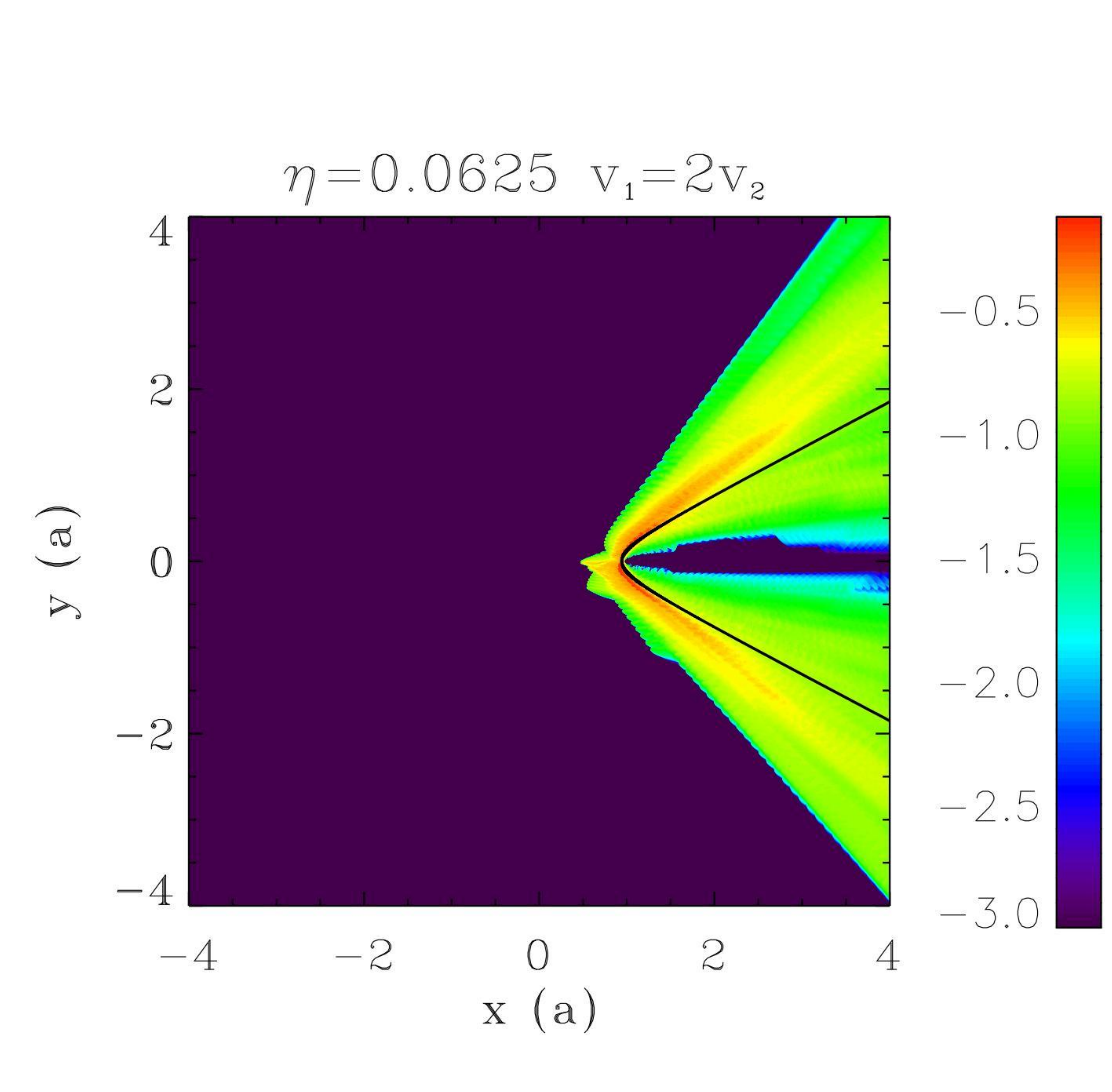}
  \caption{Top row: density maps for $\eta=1$ with $v_{1\infty}=v_{2 \infty}$ (left panel, from the same model shown in Fig.~\ref{fig:evolution}) and $v_{1\infty}=2v_{2\infty}$ (right panel). Second row: corresponding time-averaged r.m.s. of the velocity fluctuations (on a log scale). Bottom two rows: same for $\eta=1/16=0.0625$.}
  \label{fig:NTSI_KH}
\end{figure}

In real systems the velocities of the winds are never exactly equal and the contact discontinuity is subject to the KHI. Even for a $1\%$ velocity difference between the winds, this instability theoretically has a larger growth rate than the TAI and NTSI. Fig.~\ref{fig:NTSI_KH} compares simulations for $\eta=1$ with equal winds or $v_{1\infty}=2v_{2\infty}$, subject to the KHI. We also include here a map of the r.m.s. of the velocity fluctuations observed over a long averaging period. There is little difference in the outcome between equal winds and $v_{1\infty}=2v_{2\infty}$, either in the appearance of the turbulent region (top row) or in the r.m.s. of the perturbations (second row). If anything, the KHI seems to increase slightly the region where strong fluctuations occur. The NTSI dominates the final non-linear phase even when the KHI is initially present. The r.m.s. values close to one are the expected outcome of the NTSI \citep{1994ApJ...428..186V}.

We found the same results for simulations with  $\eta=1/16=0.0625$. The corresponding density maps and velocity perturbations are given in the bottom two rows of Fig.~\ref{fig:NTSI_KH}. The NTSI was studied theoretically for planar shocks but the $\eta=0.0625$ simulations show it is also present and dominant when the shock is curved, although following it requires high numerical resolutions. The simulations were performed with $n_x=128$ and 5 levels of refinement in a box of size $8a$. For lower resolutions the NTSI is not triggered and the final result is stable (the same is observed for $\eta=1$). The density maps for equal winds and $v_{1\infty}=2v_{2\infty}$ look similar. The highest velocity perturbations are at the same location but the r.m.s values are higher when an initial shear is present. We conclude that having a velocity shear in a thin shell increases the amplitude of the perturbations but does not affect much the morphology of the unstable flow, which is mostly set by the NTSI. This is consistent with \citet{1996NewA....1..235B} who  concluded from their simulations of perturbed slabs that the KHI does not strongly modify the outcome of the NTSI.

\subsubsection{Effect of increasing pressure in the stellar winds}
\begin{figure}
  \centering
  \includegraphics[width = .23\textwidth]{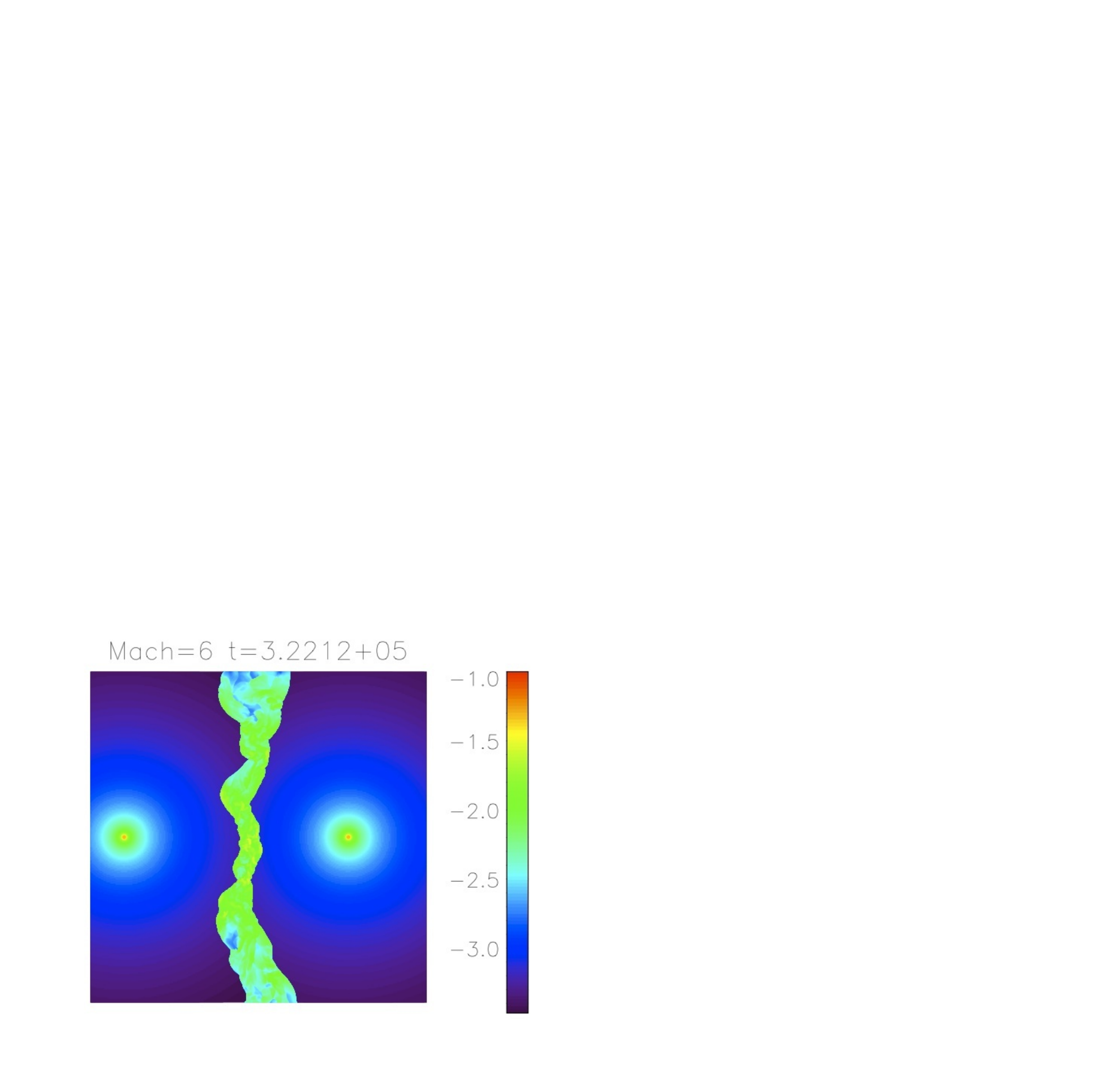}
   \includegraphics[width = .23\textwidth]{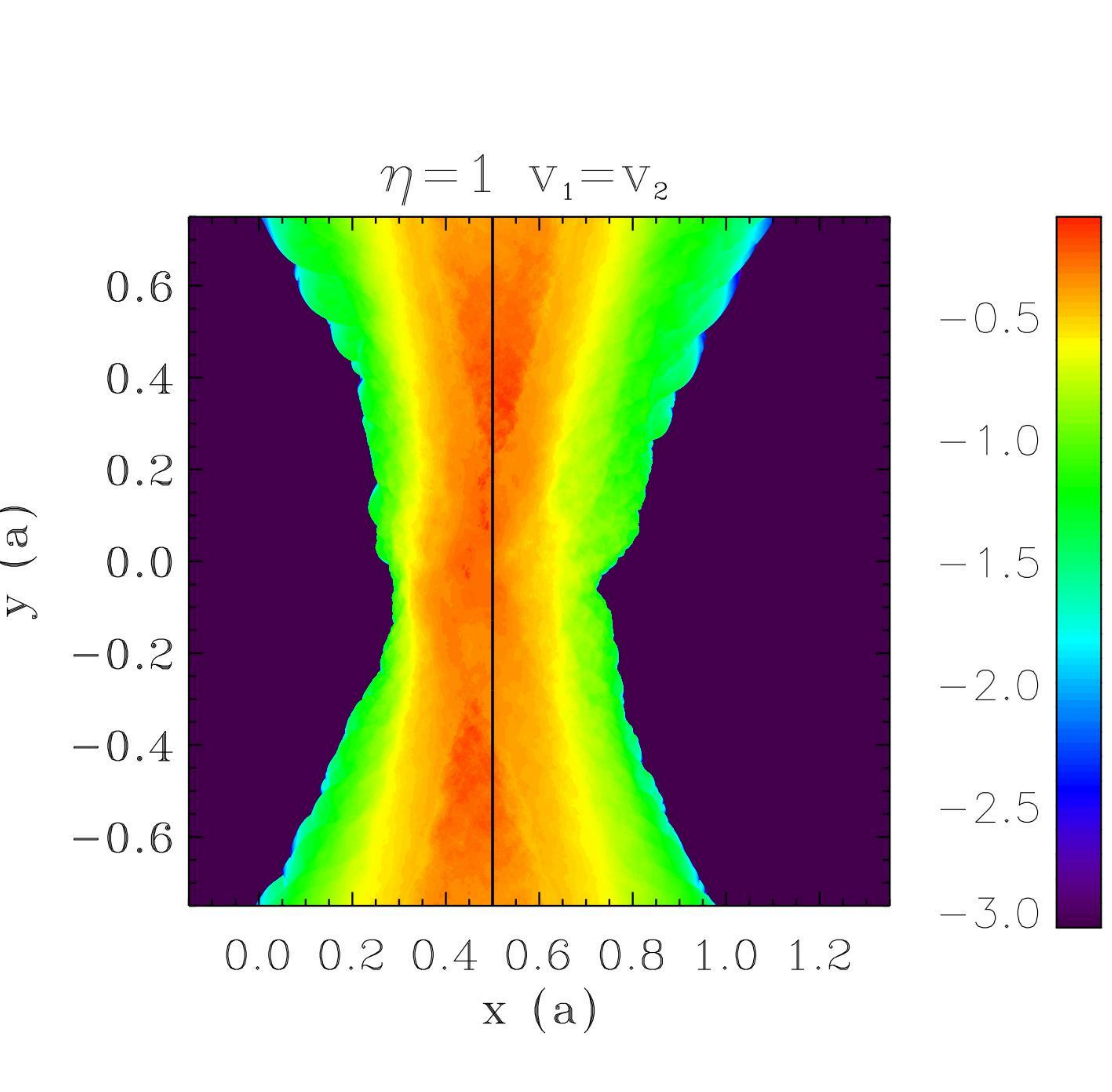}
  \caption{Left: density map of 2D colliding wing binary when $\eta=1$ , $\gamma=1.01$ and $\mathcal{M}=6$ for the highest resolution. Time is given in seconds. Right: time-averaged map of the r.m.s of the velocity fluctuations.}
  \label{fig:Mach6} 
\end{figure}

Pressure has a stabilising effect on both instabilities. We performed a simulation with $\mathcal{M}_1$=$\mathcal{M}_2$=$6$ with all other physical and numerical parameters identical to those of the $\eta=1$, $v_{1\infty}=v_{2\infty}$ simulations. Both instabilities are seen to develop but more slowly. Keeping the wind velocity constant, a lower Mach number implies a higher sound speed but the thickness of the shell increases faster so that the growth timescale  of the NTSI ($\propto L/c_s\propto 1/{\cal M}$) is longer. The NTSI is also harder to trigger as it requires a perturbation of amplitude comparable to the size of the shell. The TAI develops more slowly as pressure suppresses the development of small wavelengths perturbations in the radial directions~\citep{1993A&A...267..155D}.   The final  non-linear phase with high amplitude perturbations, shown in Fig.~\ref{fig:Mach6}, appears later than in Fig.~\ref{fig:evolution}. The shell is indeed thicker and presents smaller density contrasts than for high Mach numbers. Comparing Fig.~\ref{fig:NTSI_KH} with  Fig.~\ref{fig:Mach6}, the amplitude of the variations in shock location or the r.m.s of the fluctuations do not appear to change much but the oscillations in shock location seem to have a longer wavelength. 

\subsection{A comparison of unstable adiabatic and isothermal cases}
Finally, we compare the non-linear outcome of simulations with unstable colliding wind regions in the isothermal and adiabatic cases.  Figs.~\ref{fig:KH_eta1}-\ref{fig:KH_eta16} and Fig.~\ref{fig:NTSI_KH} show cases with $\eta=1$ or $\eta=1/16$ and $v_{1\infty}=2v_{2 \infty}$ for both the adiabatic and isothermal cases. The r.m.s. amplitude is larger for isothermal winds than for adiabatic winds when the same wind parameters are used. The unstable region extends beyond the wings of the contact discontinuity in the case of isothermal winds, unlike the adiabatic case where most of the fluctuations seem to take place within the shocked region of the weaker wind. The NTSI creates more small scale structures and  higher density contrasts are possible when the winds are isothermal. The weaker wind still propagates freely over a significant fraction of the domain despite the strong perturbations at the interface in the isothermal case. In contrast, the adiabatic simulations show that the free flowing weaker wind is confined to a very small region (\S\ref{KH}). The wind is still expected to be confined at some distance from the star in the isothermal case (see \S3.2) but this  happens further away than in the adiabatic case even when the thin shell instabilities develop.

\section{Discussion}

\subsection{Morphology of the interaction region}
We have carried out 2D and 3D hydrodynamical simulations of colliding winds to study the morphology of the interaction region and the instabilities that can affect it when orbital motion can be neglected. We first examined the relevance of widely-used analytical estimates. The position of the standoff point is very well predicted by the standard ram pressure balance on the line-of-centres. Away from the binary axis, when $\eta$ is close to 1, the opening angle of the contact discontinuity is well approximated by the analytical solution proposed by  \citet{Canto:1996jj}, which assumes conservation of mass and momentum in a thin shell. The semi-analytical solution of \citet{Stevens:1992on},  which assumes balance of the ram pressures normal to the surface,  is a better approximation when $\eta\ll 1$. This clarifies the range of validity for these approximations that have found widespread practical use in the literature. 

Numerical simulations also show that the weaker wind can be fully confined for low $\eta$, with the presence of a backward termination (reconfinement) shock, for both isothermal and adiabatic winds. The region where the weaker wind propagates freely is reduced when the Mach number of the wind is small, when the KHI develops or when the wind is isothermal. This may have some observational consequences.  One possibility is that the lines from the confined wind show unusual profiles or intensities because the wind terminates very close to the star. Another possibility is stronger, variable absorption instead of smooth absorption when the line-of-sight crosses the region where a freely-expanding wind is expected. 

More realistic simulations would include wind acceleration and radiative inhibition or braking \citep{1994MNRAS.269..226S,1995ApJ...454L.145O,2009MNRAS.396.1743P,2011arXiv1104.2383P}. The wind velocity at the stagnation point is then different from its asymptotic value, increasingly so when ram pressure balance occurs close to one of the stars. The principal consequence is to change the location of the stagnation point \citep{Antokhin:2004hi}. The basic geometry of the interaction region does not change although the asymptotic values e.g. of the contact discontinuity are probably best described by some effective $\eta$. In some extreme cases a stable balance may not be achieved and the wind-wind collision region collapses onto the star with the weaker wind \citep{Stevens:1992on,1998MNRAS.300..479P}. Another possible consequence is that a velocity shear may appear even if the coasting velocities of the winds are assumed to be equal, generating the KHI where it would not be expected. 

Orbital motion must be included when studying the large-scale structure of colliding winds.  The interaction region wraps around the binary at distances of order $v_{\infty}P_{\rm orb}$, where $v_{\infty}$ is the velocity of the stronger wind  \citep{1999IAUS..193..298W}. On smaller scales (intra-binary), a non-zero orbital velocity skews the interaction region by an angle $\tan \alpha \sim v_{\rm orb}/v_\infty$ at the apex \citep{2008MNRAS.388.1047P}. The opening angles of the shocks are slightly modified on the leading and trailing edges but the morphology of the interaction region does not dramatically change on scales $\ll v_{\infty}P_{\rm orb}$ \citep{Lemaster:2007sl}. Exploratory simulations show that the reconfinement shock is still present when orbital motion is included in a low $\eta$ model. According to our results (\S\ref{3dstudy}), no such shock is expected to form in the adiabatic simulation of \citet{2011A&A...527A...3V} since it has $\eta=1/7.5\approx0.14$. Reconfinement shocks can occur at some phases and not at others in binaries with highly eccentric orbits, as different cooling or wind velocities  are probed when the separation changes (e.g. the periastron passage of the $\eta$ Carinae, see \citealt{2011ApJ...726..105P}). The morphology also depends on the history of the shocked gas and can exhibit strong hysteresis effects in eccentric systems \citep{2009MNRAS.396.1743P}.

\subsection{Impact of instabilities}
Hydrodynamical instabilities have a major impact on the structure of the colliding wind binary. Although the overall aspect of the interaction region can still be recognised in a time-averaged sense, the wind interface can become highly turbulent, generating strong time and location-dependent fluctuations in the flow quantities. Velocity shear at the contact discontinuity in the shock region leads to the development of the KHI. An accurate Riemann solver is required to follow this instability. Eddies are already present at the interface even with a 10$\%$ velocity difference. The amplitude of the perturbations can be significant with r.m.s. values in the tens of percent for the case of adiabatic colliding winds with $v_{1 \infty}=2v_{2 \infty}$. The mixing is limited to the region of the weaker wind, with the strongest perturbations located close to the initial contact discontinuity.  The KHI has no impact on the location of the stagnation point. Equal winds are not expected to trigger the instability but introducing orbital motion was found to generate a small velocity shear even for this case \citep{Lemaster:2007sl}. Curiously, \citet{2011A&A...527A...3V} find the opposite {\em i.e.}  no KHI for nearly adiabatic winds with orbital motion, $v_{1 \infty}=1.3 v_{2 \infty}$ and $\eta=0.14$. We would expect to see significant mixing in the inner binary system, where the interaction region is only slightly skewed, unless it is dampened by numerical diffusion.

In isothermal simulations, an instability reminiscent of the TAI develops initially away from the binary axis. A second instability develops on the axis whose growth rate and properties identify as the NTSI. The NTSI dominates the non-linear evolution of isothermal colliding winds, leading to highly turbulent structures and large amplitude fluctuations in the location of the interface, including the stagnation point on the binary axis. Our results confirm the conclusions of \citet{1998NewA....3..571B} who stressed the dominance of the NTSI and the stabilising effect of pressure in their simulations of bow shocks. They also saw `wiggles' developing early on in the shock with the same properties as those we attribute to the TAI-like instability. The trigger for the NTSI is not discussed but it is likely provided by the wiggles. However, they did not attribute these to the TAI and instead argued that the TAI acts only once the shell is perturbed by the NTSI. 

The presence of instabilities in real systems is probably unavoidable. The KHI may lead to  moderate mixing of the material in adiabatic situations. The strongest mixing is obtained for high velocity shears which, in astrophysical systems, is likely to mean that at least one of the winds is radiatively efficient and not adiabatic. The radiative efficiency of the wind is classically parametrized by the ratio $\chi$ of the cooling and advection timescales, which can be evaluated as  \citep{Stevens:1992on}
\begin{equation}
\chi\approx \left(\frac{v}{1000~{\rm km}\, {\rm s}^{-1}}\right)^4\left(\frac{a}{10^{12}\, {\rm cm}}\right)\left(\frac{10^{-7}\, {\rm M}_\odot\, {\rm yr}^{-1} }{\dot{M}}\right)
\end{equation}
with $\chi\ga 3$ for an adiabatic wind and $\chi\la 3$ for a radiatively efficient wind. The ratio $\chi_1/\chi_2$ is therefore $\propto  (v_1/v_2)^5 \eta$. Because $v$ appears with a large power, a significant difference in wind velocities essentially implies that the slowest wind will be close to isothermal. In this case, thin shell instabilities develop but their outcome may be different because of the stabilising effect of thermal pressure from the neighbouring adiabatic shock \citep{Stevens:1992on,1998Ap&SS.260..215W,2009MNRAS.396.1743P,2010MNRAS.406.2373P,2011A&A...527A...3V}. For thin, highly radiative shocks, the NTSI can probably be triggered by wind variability or changes in shock width as $\chi$ varies along the orbit, if it is not already triggered by the TAI or KHI. The saturation amplitude depends strongly on the radiative losses and including a realistic cooling function in the energy equation of the fluid is essential for a detailed comparison with observations (\citealt{1995ApJ...449..727S,1996A&A...315..265W}). The shock will necessarily be larger than the idealised isothermal case so the saturation amplitudes of the fluctuations can be expected to be in between the adiabatic and isothermal cases. Other instabilities may also be at work in radiative shells \citep{1982ApJ...261..543C,1996A&A...315..265W}. Compressed magnetic fields in the shock region, if present, can also modify the growth rates and saturation amplitudes. For instance, the KHI is stabilised when the flow is parallel to the magnetic field and the velocity shear is smaller than the Alv{\'e}n speed \citep{1968RvMP...40..652G}. \citet{2007ApJ...665..445H} find that an ordered magnetic field has a stabilising effect on the NTSI in a thin slab.

In conclusion, the impact of the instabilities studied here is conveniently summarised by saying that some amount  of variability and mixing is expected in all cases but that the strongest variability and mixing are expected to be  associated with the most radiative (hence luminous) colliding winds.

\subsection{Computational requirements}
Following these instabilities is computationally demanding, especially for low momentum flux ratios $\eta$, and imposes a minimum spatial resolution together with an accurate Riemann solver. There are three numerical constraints on the spatial resolution. First, there must be enough cells within the stellar masks to properly generate the winds. For a coasting wind the mask can be larger than the actual size of the star. This cannot be the case if the stagnation point is close to one of the stars low $\eta$) and/or if wind acceleration, braking or inhibition is taken into account. The second condition is that the resolution must be sufficient to resolve the location of the stagnation point on the binary axis. This is increasingly demanding as $\eta$ decreases, but the increase in computational cost is steeper when working in the 2D setup (see \S\ref{analytical}). The last conditions relates directly to the instabilities. 
For $\eta=1/32=0.03125$, in a $8a$ simulation box, we found that a simulation with $n_x=128$ needs 7 levels of refinement in order to avoid numerical damping of the instabilities. At lower resolutions we see the initial development of the TAI far from the binary but it is quickly advected out of the simulation box without being maintained. The NTSI is not triggered and the final result is stable. We find that the shell needs to be resolved by at least 4 computational cells on the binary axis in order to develop the NTSI. Resolving the shell {\em i.e.} shock structure is the stringiest constraint on the numerical resolution. The thickness of the shell for the 2D adiabatic simulations given in Fig.~\ref{fig:shock_pos} (upper left panel) can be used to estimate the numerical resolution required to achieve this for a given $\eta$. It drastically decreases for low values of $\eta$ (slightly less so in 3D, which show thicker structures when $\eta\leq 1/32=0.03125$, see \S\ref{3dstudy}). The shell width is thinner in the isothermal case so the values derived from Fig.~\ref{fig:shock_pos} are strict lower limits for the required resolution.

Large scale simulation of a system with low $\eta$ and isothermal winds require high resolutions for the instabilities to develop. The NTSI develops at slightly lower resolutions when the KHI is present and acts as the initial seed perturbation. For instance, with $\eta=1/32$, isothermal winds and $v_{1\infty}=2v_{\infty}$ the NTSI develops with 6 levels of refinement instead of 7 in the case of equal winds.  However, it seems that the effect decreases with lower values of $\eta$. The shell always needs to be resolved, if only minimally, because the NTSI involves an imbalance of momentum {\em within} the thin shock layer. The Kelvin-Helmholtz instability in adiabatic winds is easier to model. It develops even for low resolution simulations when the velocity difference between both winds is large enough. For $\eta=1/32$, adiabatic winds and $v_{1\infty}=2v_{\infty}$ the instability develops for 4 levels of refinement. The study of the large scale 3D evolution of unstable colliding winds remains a tremendous computational challenge.

\section{Conclusion}
We have studied the morphology and the instability of colliding wind regions using numerical simulations. Compared to previous works, our study extends to much lower values of the wind momentum ratio, larger simulation domain and higher spatial resolution thanks to adaptive mesh refinement. We investigate the applicability of semi-analytical estimates for the contact discontinuity, finding that the solution of \citet{Stevens:1992on} is the best approximation to the asymptotic opening angle for small $\eta$. We find that the weaker wind can be entirely confined to a small region instead of expanding freely up to infinity over some solid angle when low $\eta$ colliding winds are considered in both the isothermal and adiabatic limits. Instabilities in the colliding wind region are important because of the mixing and variability they induce. Resolving the shock structure is required to follow the development of instabilities, which imposes increasingly stringent minimal numerical requirements for smaller $\eta$. Simulations that do not meet these requirements artificially dampen the instabilities that may be present. We follow the evolution of the KHI triggered by the velocity shear at the contact discontinuity between two winds and show that the eddies yield large fluctuations even for moderate initial shears. We formally identify the NTSI in our isothermal simulations and find that it dominates the long-term behaviour. Another instability, similar to the TAI, is present at the beginning of the simulations. Thin shell instabilities yield  large fluctuations of the flow quantities over a wide region. Our study clarifies several issues in colliding wind binary models and provides a basic framework to which the results of more complex simulations, including additional physical effects, can be compared.
\section*{Acknowledgments}
We thank Geoffroy Lesur for discussions and remarks that helped improve this study. AL and GD are supported by the European Community via contract ERC-StG-200911. Calculations have been performed at CEA on
the DAPHPC cluster and using HPC resources from GENCI- [CINES] (Grant 2010046891).

\bibliographystyle{mn2e}
\bibliography{cwinst}

\end{document}